# Oncology and mechanics: landmark studies and promising clinical applications

Stéphane Urcun[a,b,c], Guillermo Lorenzo[d,e], Davide Baroli[f,g], Pierre-Yves Rohan[b], Giuseppe Sciumè[c], Wafa Skalli[b], Vincent Lubrano[h], Stéphane P.A. Bordas[a,*]

[a]*Institute for Computational Engineering Sciences, Department of Engineering Sciences, Faculté des Sciences, de la Technologie et de Médecine, Université du Luxembourg, Campus Kirchberg, Luxembourg*
[b]*Institut de Biomécanique Humaine Georges Charpak, Paris*
[c]*Université de Bordeaux, France*
[d]*Oden Institute for Computational Engineering and Sciences, The University of Texas at Austin, Austin*
[e]*Department of Civil Engineering and Architecture, University of Pavia*
[f]*Università della Svizzera Italiana, Euler Institute, Lugano*
[g]*Aachen Institute for Advanced Study in Computational Engineering Science, Rheinisch-Westfälische Technische Hochschule Aachen, Aachen*
[h]*Hôpital Pierre-Paul Riquet, Toulouse*

**Abstract**

Clinical management of cancer has continuously evolved for several decades. Biochemical, molecular and genomics approaches have brought and still bring numerous insights into cancerous diseases. It is now accepted that some phenomena, allowed by favorable biological conditions, emerge *via* mechanical signaling at the cellular scale and *via* mechanical forces at the macroscale. Mechanical phenomena in cancer have been studied in-depth over the last decades, and their clinical applications are starting to be understood. If numerous models and experimental setups have been proposed, only a few have led to clinical applications. The objective of this contribution is to propose to review a large scope of mechanical findings which have consequences on the clinical management of cancer. This review is mainly addressed to doctoral candidates in mechanics and applied mathematics who are faced with the challenge of the mechanics-based modeling of cancer with the aim of clinical applications. We show that the collaboration of the biological and mechanical approaches has led to promising advances in terms of modeling, experimental design and therapeutic targets. Additionally, a specific focus is brought on imaging-informed mechanics-based models, which we believe can further the development of new therapeutic targets and the advent of personalized medicine. We study in detail several successful workflows on patient-specific targeted therapies based on mechanistic modeling.

*Corresponding author
 Email address:* `stephane.bordas@alum.northwestern.edu` (Stéphane P.A. Bordas)

**Introduction**

Cancer is one name for numerous diseases. Each location leads to a unique physiological framework; within each location, different forms of cancer are identified, and this diversity is subdivided toward patient specific cases. Additionally to individual diversity, cancer should be considered as a multiscale disease at the cellular and sub-cellular scales, where the micro-environment recruitment by tumor cells and genetic instabilities play an equally important role [1, 2].

Despite the challenges posed by the heterogeneous and multiscale nature of cancer diseases, the recent decades have witnessed much progress in their clinical management, as shown in the global cancer statistics for 2020 [3]. Firstly, there has been a reduction in the incidence thanks to the prevention of risk factors present in the lifestyle of transitioned countries, such as tobacco, alcohol and obesity. These risk factors participate in the most common cancers of the lung, colorectal, liver, stomach and bladder. Cancer incidence has also been reduced by treatment of biological causes like bacterial and virus infection, hormone exposure, poor hygiene or flawed food storage. Before these treatments, these factors contributed to cancers of the lower stomach, bladder, liver, breast and cervix. Secondly, cancer mortality has also decreased thanks to the multiple advances in clinical management of tumors, such as the early detection of prostate and breast cancers, treatment breakthroughs in breast cancer and metastatic melanoma, and active surveillance in various cancers (for instance, lung, breast and thyroid). In addition to prevention and biological management, genetics has brought considerable insights and effects on treatment outcomes in specific locations (*e.g.* breast cancer stem cells [4], lung biomarkers [5]). Thanks to the recent advances in the understanding and treatment of cancer, many recent battles have been won against cancer. Childhood cancer (mostly leukemia) mortality rates have been halved in thirty years despite an increasing incidence [6]. Cervical cancer is now considered almost 100% preventable, thanks to the human papillomavirus vaccine and population screening [3]. Early detection of prostate cancer has lowered the mortality by 21% in ten years [7]. Thanks to early detection and breakthroughs in the treatment of breast cancer (inhibition of human epidermal growth factor receptor 2 (HER2) by monoclonal antibodies [8]), 5-year survival reaches 90% - all stages*[1] averaged - since 2014 in high-income countries.

In spite of this immense progress, cancer remains an important burden for societies worldwide. The changing lifestyle of economically transitioning countries will increase their relative cancer incidence. Some cancers still have a poor prognosis* (pancreas) or their etiology* is poorly understood (prostate or thyroid), with the exception of environmental and genetic risk factors. The authors of [3] predicted that by 2025 the worldwide incidence of pancreatic cancer will be higher than that of breast cancer. There is currently no effective treatment, and pancreatic adenocarcinoma* has a 5-year survival rate of only 9% [9]. Strongly correlated with socio-economic development, the incidence of colorectal cancer will rise in the economically transitioning countries within the next decades, and probably cause important damage. In economically transitioned countries, where screening and treatments are in the state of the art, the 5-year survival rate is near 90% for benign stage* I, but below 10% for metastatic stage* IV [10].

---

[1]all the words marked with a * are in a glossary at the end of this review

The biological approach, in its broad sense including biochemistry, molecular, genomics and phenotypic* aspects, has brought and still brings numerous insights into cancerous diseases, which have contributed to the aforementioned advances in the clinical management of these pathologies. The mechanical approach, at the tissue and cellular levels, includes solid, fluid, porous medium and electrophysics show effective ways to clinical applications in oncology. In this review, we understand mechanics by its two common definitions: *the science of systems* which are subjected to the fundamental principles of Newtonian mechanics and *the science of building machine*, as the design of experimental setups is from end to end a critical part of the clinical applications we present.

From the experimental side, following a biological approach and ignoring structural mechanical effects could conceal some phenomena. For instance, the wound healing, chronic fibrosis and cancer progression 'triad', comprehensively reviewed by Rybinski *et al.* in [11], implies that stroma* mechanics are a fundamental part of many cancers progression. The mechanical phenomena, implied in stroma* mechano-biology, cannot be reproduced in classical Petri dish 2D cultures [12]. The structural effects of these interplay emerge only in 3D cell cultures, as multi-cellular tumor spheroids [13]. Several phenomena, allowed by favorable biological conditions (genetic, immune or environmental), emerge *via* mechanical signaling at the cell's scale and *via* mechanical forces at the macroscale: epithelial* to mesenchymal* transition (EMT) [14, 15], metastatic cell advection by lymphatic and blood fluxes [16], or tumor cell growth inhibition and necrosis* induced by mechanical stress [13, 17]. Among complementary approaches of cancer description, mechanics is now well accepted. For four decades, mechanical phenomena that emerge from cancer have been widely studied: in 2014 a seminal review was produced by Jain *et al.* [18] for the meso and macroscale, and at the cell scale, a recent review has been provided by Northcott *et al.* [19]. In 2020, the review of Jain *et al.* has been augmented by Nia *et al.* [20] to encompass several new multiscale phenomena. This long-term research has matured enough to provide clinical applications and some inspiring examples are given in section 3.

It is now established that mechanical stress influences tumor growth, as shown by a range of sources from early *in vitro* studies [21] in 1997 to encapsulated multi-cellular spheroids [13]. This fundamental mechanism is explained in section 1.1. Computational fluid dynamics (CFD) describes viscous flow in compressible vessels [22] and is the basis of the design of microfluidic devices to isolate metastatic cells in blood samples [16]. Already effective clinical applications of CFD are presented section 3.2. Stroma* mechano-biology, focusing on the extra-cellular matrix* (ECM), fibroblasts* and their interplay with tumor cells, has been an active research field for several years (reviewed in [19], *in vitro* [23], [24], [25], and *in vivo* [15]). These investigations, presented section 1.2, enable the design of new targeted drugs, such as focal adhesion* kinases (FAKs) inhibitor [26], and a new clinical prognosis* tool [15], which are presented section 3.1. The more mature mechanical field with experimental devices applied to oncology in the physics of electricity. In this review, we will mainly focus on the non-thermal irreversible electroporation. The reader will find the history of this field in the book chapter of Rolong *et al.* [27]. Since the first landmark study in animal models, published in 1957 [28], irreversible electroporation has evolved to a well-established clinical practice for unresectable tumors. Recent developments led to the coupling of irreversible electroporation and immunotherapy with promising clinical applications in pancreatic cancer [29]. Section 3.3 of clinical applications of electrophysics is complemented by the presentation of a new paradigm: tumor-treating fields, which open promising opportunities in the standard of care of glioblastoma* multiforme.

Mechanical-based modeling considers the cancer dynamics as a physical system subjected to the fundamental principles, as balance laws. In solid mechanics, the *in vitro* findings of tumor growth inhibition now play a part in patient-specific modeling [30, 31, 32]. As a wide majority of living tissue may be modeled as a composite system of fluids and solids, with a variable permeability, they can be described by poromechanics. This strong coupling of a solid network of fibers saturated by fluids can describe permeable tissue, such as pancreas, liver, brain or lymph node. In this modeling framework, the infiltration of tumor cells or the diffusion of drugs within the tissue can be explicitly modeled *in vitro* [33, 34] and *in vivo* [35, 36, 37]. This emerging framework is presented in section 1.3. The interested reader will find in the supporting information (section 5) a brief presentation of the constitutive equations of the tumor growth, its different inhibitions and specificity of poromechanics.

The progress in the understanding of the multiscale mechanisms, and the numerous hypotheses of mathematical models, have led to a consequent increase of potential therapeutic targets. But these advances created new problems: comprehensive clinical trials or patient adapted therapeutic scenario become virtually impossible, the number of parameters to control being overwhelming. Thus, the necessity appears of a systemic *in silico* exploration of therapeutic targets and patient scenarii against clinical evidence. To address this challenge, tumor forecasting has emerged as a viable option [38]. This strategy combines the integration of patient-specific standard clinical and imaging data with biomechanical models to predict the growth and treatment response for individual patients via computer simulations. Indeed, this approach can be compared to the well-established methods of weather forecasting. Since the 2010s, progress in medical imaging has enabled a wide range of translational advances for biomechanical modeling of the growth and therapeutic response of tumors. As claimed in the seminal paper of Yankeelov *et al.* in 2013 [39]: "Magnetic resonance imaging (MRI) and positron emission tomography (PET) have matured to the point where they offer patient-specific measures of tumor status at the physiological, cellular, and molecular levels". Patient-specific clinical data can give information at the gene scale, the micro-scale (cells and capillaries) and the macro-scale. Hence multi-scale models can be initialized by clinical data.

The association of imaging-informed modeling and computational mechanics brings notable advantages on a patient-specific basis, for instance a 3D MRI can be converted into a computational domain, and its segmentation into subdomains. Multi-parametric MRI allows for the monitoring and quantification of many properties, such as fluid proportion, cellular density, cell metabolism and, using dynamic contrast enhanced MRI (DCE-MRI), dynamic physical quantities such as the diffusion of chemical species and the vascular permeability. Since a 3D MRI can be converted into a computational domain, the patient-specific forecast can be directly compared against further clinical imaging data. The clinical imaging methods that can be used within this framework are presented section 2.1. Imaging-informed models of cancer growth and therapeutic response can be directly used in the clinic, end-to-end. Model validation can also be assessed using a combination of the same types of imaging and clinical data used for initialization and parameter identification, but collected at one or multiple posterior dates during the course of tumor monitoring or treatment (for instance, see [31, 40]). Detailed workflows of imaging-informed mechanical-based are provided section 2.2.

This review specifically aims to present the process by which mechanics can step in clinical management of cancer. We divided the review into three topics:

1. Pre-clinical studies, *in vitro*, *in silico* or in an animal model. Mechanical phenomena or mechanical therapeutic targets can be tested in isolation from their complex micro-environment. These studies also facilitate the collection of wealth of data to calibrate and validate biomechanical models describing such phenomena and therapies;
2. Patient-specific tumor forecasting. Computational mechanics can be used as a diagnostic, staging[*] and prognostic tool. This section is specifically designed for students;
3. Clinical applications on cancer management directly derived from mechanical findings in cellular mechano-biology, fluid mechanics, and electrophysics.

For each topic, we recommend further specific reviews for the interested reader. At the beginning of each section, we present the major findings, methods and remaining challenges. At the end of each section, we provide a set of short statements to summarize the topic. The review ends with supplementary information in section 5. It contains an introduction to the different modeling frameworks, a glossary of clinical terms, and a list of acronyms.

**1. Pre-clinical studies**

*In vitro*, *in silico* and animal models studies share the same goal of reducing the complexity of cancer phenomena. These simplifications naturally limit the scope of these studies. Nevertheless, some of them have led to new therapeutic targets, which are through clinical trials. We present three facets of mechanical understandings provided by pre-clinical studies: mechanical inhibition of the tumor growth, stroma* mechano-biology and poromechanical modeling of tumor tissue. For each subject, we selected a few studies which clearly expose the mechanical phenomena at play.

*1.1. Growth inhibited by mechanical stress*

In ancient Greece, the palpation of stiffened tumor tissue was one of the only accessible diagnosis tools. This abnormal stiffening is one of the results of mechanical stress concentration. In 1997, the *in vitro* study of Helmlinger *et al.* [21] unveiled part of these mechanisms: high mechanical stress inhibits tumor growth. Elements of the modeling of this phenomenon are briefly presented in the supplementary information (section 5.1). Nowadays, these results could have consequences on treatment management of solid tumors in clinics, such as prostate [30] or breast [31] cancers (see section 2.2). The relationships between mechanical stress and tumor growth are complex and non-linear. If a cell line is subjected to a mechanical load considerably higher than its usual osmotic pressure, it will almost certainly lead to a growth inhibition. We here select few examples, each with its specific mechanism:

- non-hypoxic* driven apoptosis* by Cheng *et al.* in 2009 [17] (*in vitro*)
- smooth confinement by spheroid encapsulation by Alessandri *et al.* in 2013 [13] (*in vitro*)
- growth inhibition applied as treatment before surgery by Brossel *et al.* in 2016 [41] (in animal models)

The interested reader could peruse the sections devoted to this topic in the reviews of Jain *et al.* in [18] and Nia *et al.* in [20].

*Non-hypoxic\* driven apoptosis\**. Spheroid growth inhibition in agarose gel is a well-known *in vitro* experimental protocol, enhanced by Cheng *et al.* in 2009 [17]. The tumor spheroids are embedded with fluorescent micro-beads in agarose gel and the 3D distribution of micro-beads surrounding growing spheroids is recorded using confocal microscopy. The change in micro-bead density is converted to strain in the gel, thus the compressive stress around the spheroids can be estimated.

There is a strong correlation between the solid stress distribution and spheroid shape. It has been shown that if compressed, spheroids overexpress anti-apoptotic* genes [24]. However, regions of high mechanical stress induce apoptotic* cell death (proliferating and necrotic* cells in spheroids were detected with a proliferative cell fluorescence marker, see Figure 1A). The important implication if this result is that apoptosis* via the mitochondrial pathway, induced by compressive stress, may be involved in tumor dormancy, in which tumor growth is constrained by the balance of apoptosis* and proliferation.

*Smooth confinement by spheroid encapsulation.* Cellular capsule technology is an *in vitro* experimental protocol developed in [13] where tumor spheroids were cultured within spherical porous alginate capsules. The capsules are generated by co-extrusion using a 3D printed device and the alginate pore's size allows a free flow of nutrients. The capsule dilatation exhibits an elastic deformation with negligible plasticity and no hysteresis. As their size and thickness can be precisely controlled and alginate gel has isotropic and incompressible properties, the internal pressure can be analytically retrieved from the capsule deformation. Hence, when the tumor spheroid comes in contact with the inner wall, the capsule works as a mechanosensor. From their deformation, one can retrieve the stress state within the tumor spheroids.

In this respect, the post-confluent growing tumor can be regarded as a tumor model that grows against the surrounding tissues and organs. A digital twin of this technology was created through multiphase poromechanical modeling by Urcun *et al.* in [42]. It showed that a poromechanical framework is capable of reproduced both the capsule evolution and its inner structure.

*Growth inhibition applied as treatment before surgery.* Gradient of magnetic field is an experimental protocol developed in [41] to show, *in vivo*, the effect of a mechanically constrained tumor growth. The human breast cancer cell line, MDA MB 231, admixed with ferric nanoparticles is grafted subcutaneously in nude mice. Two magnets located on either side of the tumor create a gradient in the magnetic field, acting on the nanoparticles. Through their reaction, the magnetic energy is transformed into mechanical energy, thereby applying a biomechanical stress to the tumor. The mice are divided into four groups. The first group is subjected to a 2-hour exposure per day to a magnetic field gradient for 21 days. The three other groups are control groups: group 2 has a tumor with nanoparticles without exposure to a magnetic field, group 3 has a tumor without nanoparticles but with the same 21 days treatment as group 1, and group 4 has a tumor without nanoparticles and without treatment. All mice are sacrificed on Day 74.

There was a significant difference of 61% between the median volume of treated tumors and untreated controls in the mice measured up to Day 74. This demonstration of the effect of stress on tumor growth *in vivo* suggests that biomechanical intervention may have a high translational potential as a therapy in locally advanced tumors (stage* II-III). To our knowledge, this experiment is not yet modeled by a mechanical system.

Mechanically-induced inhibition of tumor growth is a well-known phenomenon that has consequences not only on *in vitro* findings, but also *in vivo* modeling (see section 2.2). Moreover, the effect of mechanical stress can also be described at the cell's scale, where interaction between fibrous tissue and carcinoma* cells is a common situation. This microscale description leads to critical *in vitro* findings and to new clinical applications (see section 3.1).

*1.2. Stroma mechano-biology*

The influence of the tumor microenvironment on disease evolution has become central in cancer research [43]. We focus here on a small part - but critical from the mechanical point of view - of the microenvironment: stroma* cells, precisely fibroblasts*, and ECM. Generally speaking, fibroblasts* associated with tumor cells will lead to a modified ECM. This ECM is denser and stiffer, due to the abnormal collagen production of these fibroblasts* [44]. As the tumor grows and invades the surrounding tissue, the stroma* interplay with tumor cells dramatically changes. The increase of the mechanical stress due to the proliferation of the tumor cells and to the stiffer ECM will lead, at the cell's scale, to internal and external modifications. In other words, mechanical stress leads to a modification of the tumor cells phenotype* [24]. This phenotype*'s switch depends on the cancerous cell line, so may take various expressions. We can list several frequently encountered properties within these cell lines:

- enhanced tumor cell contractility, *i.e.* the cell's mobility is increased;

- the polarity of the tumor cell cytoskeleton is altered, which leads to 'star-shaped' cells, more adapted to a dense environment (see Figure 1B);

- inhibition of the tumor cell apoptosis*, since the mechanical stress is maintained, the programmed death of cells is stopped;

- alterations on ECM mechanical properties and architecture. The production of collagen-rich ECM by fibroblasts and matrix metalloproteinase - enzyme which degrades ECM - by cancerous cells deregulates the composition and the architecture of the ECM.

These changes are of critical importance for the disease evolution, as invasiveness and therapy resistance. They also constitute new and promising therapeutic targets, which are exposed in section 3.1. The interested reader would peruse the review of cellular mechano-biology of cancer by Northcott *et al.* in [19].

*Mechanical stress, integrin and malignant phenotype.* have been studied by Paszek *et al.* in 2005 [24]. Abnormal tissue stiffness has been historically associated with tumors and, nowadays, breast cancer patients with stiff fibrosis - an excessive activity of fibroblasts - are known to have poor prognosis* [45]. The authors of [24] lead unconfined compression tests by indentors on both sane and malignant mammalian tissues, mammalian epithelial* cells and fibroblasts*. The samples are associated to collagen gels of various stiffnesses, from 170 Pa to 1.2 kPa. This range, at its lower bound, corresponds to the common stiffness of adipose mammalian tissue and, at its upper bound, to fibrosis. The cell colonies evolve accordingly to the stiffness of their surrounding environment (for the mammalian epithelial* cell case, see Figure 1B, stiffness range from left to right). The differences between these evolution show, *in vitro*, the phenotype*'s switch described at the beginning of the section. Moreover, the authors of [24] expose the key phenomenon involved in the mechanical stress signaling. Their study emphasizes the critical role of $\beta 1$ integrin (wild-type and mutant) in the abnormal evolution in every kind of sample. Integrin is a transmembrane ECM receptor which functions as a mechanotransducer. At the outer surface of the cell's membrane, the integrin protein is bent by the mechanical compression of the external medium. This deformation provokes, in the inner cell's medium, chemical reactions which precede all the biological manifestation of abnormal growth. If the mechanical downstream signaling - the inner chemical reaction - is maintained for several days, an altered phenotype will emerge. Two main mechanisms, belonging to the cellular internal medium, are exposed:

- the Rho-dependent cytoskeletal tension. This mechanism is involved in mesenchymal* stem cell differentiation. Without the signaling of cytoskeletal proteins as integrin, a mesenchymal* stem cell could evolve into a round-shaped adipose cell. If this alternate cytoskeletal tension is activated, the stem cells will adopt a 'spread' shape characteristic of bone cells. The reader is referred to the work of McBeath *et al.* in [46] for a detailed description.

- the epidermal growth factor(EGF)-transformed epithelia. EGF signaling plays a critical role in the initiation of the epithelial* to mesenchymal* transition (EMT). EMT is the deregulation of the highly organized epithelial tissue and is characteristic of malignant carcinoma*. EGF is a transmembrane protein which is responsible for the downregulation of E-cadherin. And E-cadherin, like integrin, is transmembrane protein responsible for cell adhesion. The downregulation of E-cadherin implies an upregulation of the integrin adhesion. This upregulation will lead to the clustering of integrin on the cell membrane and deregulation of focal adhesions*. This specific phenomenon constitutes a new therapeutic target treated section 3.1. The interested reader is referred to the work of Lindsey *et al.* in [47], for the specific role of EGF in this phenomenon.

In the last part of their experiment, Paszek *et al.* shows that the inhibition of $\beta 1$ integrin or EGF signaling can revert to the malignant phenotype expressed in the cell-lines undergoing fibrosis condition. This early *in vitro* study pointed out potential therapeutic targets in the malignant evolution of carcinoma*.

Whatever the scale of interest of these phenomena, macro and micro, they share a common interplay between fluids, solids, and biochemical reactions. Poromechanics allows for modeling the chemical agent transport by the fluids, though the solids. To explicitly model this complex interplay, poromechanics emerges as particularly adapted framework.

*1.3. Reactive poromechanics*

This framework allows for explicitly modeling fibrous tissue, cells and surroundings fluids into a strongly coupled physical system. Used since the XIX$^{\text{th}}$ century in soil mechanics, poromechanics now emerges as a new framework for living tissue modeling (see *Encyclodedia of biomedical engineering*, chapter *Poroelasticity of living tissue* [48]). In poromechanical modeling, the tumor system is considered as a multiphase continuum: a solid scaffold in which and through which fluids flow. This framework is briefly presented in the supplementary information section 5.1. The flows are driven by pressure gradient, and subjected to the ratio of the solid's permeability under the dynamic viscosity of the fluids. This relationship, in its usual form, is known as Darcy's law. The solid scaffold may be considered as rigid or deformable. If it is deformable, the momentum conservation of the porous system is composed of both fluids contributions (usually isostatic pressure) and solid contribution (elastic, hyperelastic, ...). Living tissue modeling implies reactive biochemical agents, which can be produced or absorbed by both solid and fluid components, and are subjected to advection-diffusion equations. This complete system is denoted as a reactive porous medium. We highlight the potential of this framework with two pre-clinical examples:

- Modeling drug resistance in Non-Hodgkin's lymphoma*, validated by histological* cuts. Tumor inner structures, accordingly to the cell lines, are accurately modeled by Frieboes *et al.* in [35] (animal model);

- Modeling drug delivery in brain by multi-compartment porous medium. In this example, the solid scaffold is deformable. Provide several qualitative results in anisotropic diffusion and pressure interplay in brain tissue, by Ehlers and Wagner in [36] (*in silico*).

*Modeling drug resistance in Non-Hodgkin's lymphoma**. In the experiment reproduced by Frieboes *et al.* in [35] two cell lines of lymphoma* are injected into two groups of mice. The first cell line is drug-resistant, thereafter denoted p53, and the other is drug-sensitive, denoted Arf. On day 0, five histological sections, equally distributed in space, are performed in three tumors of each type, extracted from sacrificed mice. The stainings (H&E, Ki-67, HIF-1$\alpha$ among others) give cell viability, necrosis*, proliferation, apoptosis*, oxygen diffusion and blood vessel density. These quantities are used to calibrate the model parameters. On day 21, the observations are made *in vivo* by intravital microscopy.

Frieboes *et al.* build a mechanical model belonging to the hybrid category. In mechanical-based modeling, one can distinguish three categories: continuous, discrete and hybrid. Continuous - or homogenized - modeling consider the phenomena at a sufficiently large scale to hypothesize the absence of strong discontinuities within the modeled object. Conversely, the working hypotheses of discrete modeling are at the cell's scale, and each cell - also denoted agent - has its own characteristics, with possible discontinuities between the cells. While continuous modeling, by its homogenization, implies a controlled loss of information, the computational cost of discrete modeling becomes prohibitive as the number of cells grows. Hybrid modeling, a sort of multi-scale modeling strategy, attempts to make the best of both worlds.

The hybrid system, at the lymph node scale, is governed by advection-diffusion within a continuous domain. Here, this domain is a porous medium, say the tumor cell infiltration within the extra-vascular space is governed by Darcy's law. However, the system parameters are calibrated with information on the cell's scale through histological staining and intravital microscopy. This protocol was described in 2012 by Macklin *et al.* [49] and applied on ductal adenocarcinoma* in situ (a precursor of invasive breast cancer).

The following experimental results are retrieved from the model:

- both types of tumors reach the same size within the lymph node: $5.2 \pm 0.5$ mm diameter.

- their inner structures of the tumor are different, according to their cell-line. Arf cell-line tumor is denser in the peripheral region, contrary to p53. The high core density of p53 cell-line tumor could be associated with its drug resistance, by decreasing the penetration of chemical agent. The p53 cell-line has a 4-fold higher density of endothelial cells - cells constituting the wall to the blood vessels - in the core of the tumor than the Arf cell-line. The p53 proliferating cells fraction is 2-fold higher than the Arf cell-line. Hypoxia* is higher in the peripheral region of the p53 tumor. The apoptotic p53 cell density in the whole tumor is 2-fold higher than the Arf, suggesting a non-hypoxia* driven apoptosis* in the p53 cell line.

These rich results show the power of this poromechanical multi-scale approach, which could bring considerable insights into *in vivo* modeling.

*Modeling drug delivery in brain by multi-compartment porous medium.* Brain tumors, especially glioblastoma*, are still a challenge from the surgery, therapy and modeling points of view. Since 2005, the protocol presented by Stupp *et al.* in [50] is the only glioblastoma* standard of care. It prescribed maximum surgical resection followed by concomitant radio-chemo-therapy for 6 weeks and chemotherapy maintenance during 6 or 12 months. Nevertheless, the median survival is only 16 months and the 5-year survival remains at 5%. Invasive drug delivery is an additional way to decrease the tumor progression, recurrence, and resistance (for instance, see the Gliadel wafers [51]). The blood-brain barrier is not permeable enough to ensure a proper drug delivery by intravenous application, hence perfusion by catheter directly into the brain parenchyma* is needed. During resection of the tumor, the surgeon drills small holes into the skull and place infusion catheters directly into the extra-vascular space of the parenchyma*.

In surgical planning software, solutions are already proposed for the simulation of the chemical agent diffu-sion by the catheter. The authors of [36] propose to improve these simulations by taking into account the porous interplay of the brain and its heterogeneous permeability and diffusion. W. Ehlers and A. Wagner are the founders of a recent theory of porous medium [52] dedicated to this application. They also worked on model reduction for clinical applications [53]. In [36], they model the brain parenchyma* as a porous medium with two separated fluid compartments: blood and interstitial fluid.

Grey matter regions may be considered isotropic in terms of permeability and diffusion. Experimental observations of white matter tracts show anisotropy in permeability and diffusion [54]. The authors assume, as Tuch *et al.* in [55], that white matter tracts directions provide the eigenvectors of anisotropy. These directions can be recovered by the DTI MRI method (see section 2.1, paragraph *Imaging methods and physical quantities*). The imaging data of this method are leveraged and used to set, on a tensorial basis, the heterogeneous permeability and diffusion of the simulation. The other material properties come from the literature but, as indicated by the authors, they could be provided by other MRI methods.

The authors do not pretend clinical relevance but propose possible interpretations of their numerical results. They simulate 2 catheter flows: $0.3$ and $0.6\,\mathrm{ml}\cdot\mathrm{h}^{-1}$. They retrieve for the case of $0.6\,\mathrm{ml}\cdot\mathrm{h}^{-1}$, in accordance with *in vivo* data, an excess of 1.1kPa (+8mmHg) in interstitial fluid pressure, which could be life threatening. Due to the interplay between the solid scaffold and the inner fluid, they found that a higher infusion surface is obtained with a constant application rate. Finally, due to the insertion of the catheter, the permeability of the damaged tissue dramatically increases. In the simulation, this provokes back-flows along the catheter shaft and this phenomenon is also acknowledged *in vivo*. Elhers *et al.* suggest that the porous modeling of the brain will help to gain mechanical insights in surgical and therapy management of glioblastoma*.

This last example is in a pivotal position between the pre-clinical section and the patient-specific forecasting section 2. Even if it does not pretend clinical relevance, it shares common features presented section 2:

- the use of patient imaging data to set the parameters of the mathematical model

- the direct comparison of the model results with *in vivo* data

Thus, poromechanics could be a suitable framework for imaging-informed modeling presented in the next section.

Table 1: **Mechanical understandings of cancer in pre-clinical studies.** For further readings, the interested reader would refer to the reviews of Nia *et al.* [20] and, for mechano-biology specifically, Northcott Nia *et al.* [19].

| Authors | Cell line | Model | Objective |
|---|---|---|
| Cheng *et al.* 2009 [17] | non-metastatic murine mammary carcinoma* 67NR | *in vitro* | Measure of the deformation of agarose gels by tumor cells |
| Alessandri *et al.* 2013 [13] | mouse colon carcinoma* CT26 | *in vitro* | Retrieve inner pressure exerted by tumor cells on alginate capsules |
| Brossel *et al.* 2016 [41] | human breast cancer cell line MDA MB 231 | nude mice | Release internal tumor stress in animal model by ferromagnetic particles as bioactuators |
| Frieboes *et al.* 2013 [35] | lymphoma* E$\mu$-myc Arf-/- and E$\mu$-myc p53-/- | mice | Forecast inner structure of tumor cell line drug resistance or not in animal model by poromechanical modeling |
| Elhers *et al.* 2015 [36] | not specified (patient) | *in silico* | Contribute to the modeling of anisotropic drug diffusion within human brain by poromechanics |
| Paszek *et al.* 2005 [24] | mammalian tissue of MMTV-Her2/neu, Myc, and Ras transgenic mice ; nonmalignant MCF10A ; HMT3522 S-1 MECs | *in vitro* | Understanding mechanical signaling involved in tumor cell phenotype switch |

**Pre-clinical studies: key points**

- The stripped-down *in vitro* environments allow for clearly exposing mechanical phenomena in cancerous disease: mechanically-inhibited tumor growth, mechanically-induced phenotype*'s switch.

- These findings led to clinical applications [30], prognosis* tools [15], and new therapeutic targets [56].

- Poromechanics is an emerging framework in the mechanistic description of cancer and could lead to new and rich physical insights.

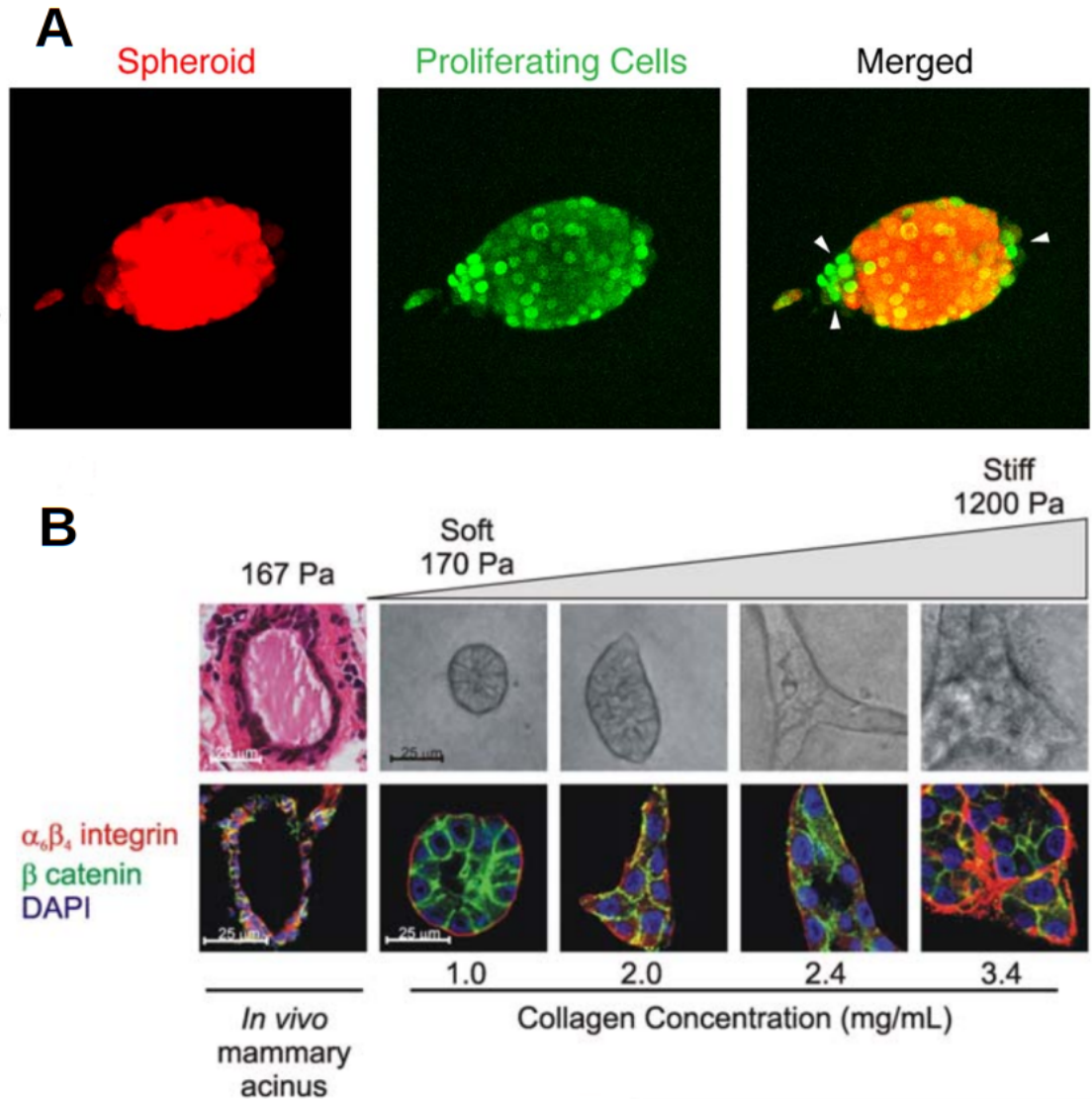

Figure 1: ***In vitro* mechano-biology of tumor growth, landmark studies.** The two panels illustrate the complex relationship between mechanical stress and tumor growth. If a cell line is subjected to a mechanical stress considerably higher than its usual osmotic pressure, it will lead to a growth inhibition. However, for some cell lines, it exists a window where a high mechanical stress will lead to phenotype alterations and a more aggressive behavior **Panel A** Mechanical inhibition of tumor growth by Cheng *et al.* in 2009 [17]. Cancer cell proliferation (green) in tumor spheroids (red) is suppressed in the direction of higher mechanical stress (*i.e.*, in the direction of the minor axis of oblate spheroids). On day 30, the compressive stress is estimated at $3.8\,\text{kPa}$. Arrow heads indicate the regions with more cell proliferation. Scale bar $= 50\,\mu\text{m}$. **Panel B** Mechanically induced phenotype switch: matrix rigidity regulates growth, morphogenesis, and integrin adhesions in Paszek *et al.* in 2005 [24]. Top: phase images and H&E (hematoxylin and eosin) stained tissue showing typical morphology of a mammary gland duct in a compliant gland ($167\,\text{Pa}$), compared with mammalian epithelial* cell (MEC) colonies grown in collagen (BM/COL I) gels of increasing stiffness ($170 - 1200\,\text{Pa}$). Bottom: confocal immunofluorescence images of tissue section of a mammary duct and cryosections of MEC colonies grown as above, stained for $\beta$-catenin (green), $\alpha 6$ or $\beta 4$ integrin (red), and nuclei (blue). The increasing size of the colonies is directly related to the increasing stiffness of the collagen.

## 2. Tumor growth forecasting: imaging-informed patient-specific modeling

This modeling paradigm was first developed by Swanson *et al.* in [57] in 2002, and after in [58, 59]. In 2013, with the progress of imaging methods, this framework was further extended by many groups (see the review of Mang *et al.* in [60]), and especially by Yankeelov *et al.* [39]. We start with a brief presentation of the clinical imaging methods that can be used to set the value of the parameters of the mathematical models. These parameters can define the material properties (diffusion, permeability, stiffness) or the material component (fluid, solid, cellular density). Moreover, the clinical imaging data provide patient-specific geometry [39], and this is a crucial gain for the relevance of the model results. However, the challenges involved in the uncertainty quantification and the use of these imaging methods in certain clinical settings may reduce their extensive use across different hospitals. Likewise, the adaptability of the mathematical models to these imaging data sources is still under investigation [61].

The second part of this section is devoted to detailed presentation of three clinically-relevant applications. In a few words, these models describe tumor growth and therapeutic response by means of reaction-diffusion equations that are constrained by the mechanical stress field induced by the development of the tumor. While they share the same modeling core, each model is adapted to the specificity of the cancer location.

### *2.1. Clinical imaging methods*

*Imaging methods and physical quantities.* Currently, the imaging methods performed in clinical oncology for diagnosis, monitoring, and assessment of treatment efficacy are:

- Computed tomography (CT) scans, used for cancer early diagnosis. CT scans results from the mathematical reconstruction of multiple axial X-ray images [62]. This reconstructed image of a large part of a body can be created in a few seconds, allowing for cancer detection throughout the body. Apart from bone material, the contrast of the various soft tissue can be enhanced by a radio-contrast agent like iodine or barium.

- Positron emission tomography (PET) scans, which consist of an intravenous injection of a positron-emitting radiopharmaceutical, like $^{18}$fluor-fluorodeoxyglucose (F-FDG). The compound is partially metabolized by cells such that it accumulates mostly in cells with the higher metabolic rate, this accumulation being then detected [38]. Contrary to CT scans, 3D images are directly obtained in a PET scan, but the length of diffusion of the degraded compound implies a maximum of 2 millimeters precision [63]. When a precise measurement is required, a PET scan is complemented by other methods.

- Magnetic resonance imaging (MRI), the most commonly used technique for cancer diagnosis and screening. The various MRI methods are capable of providing information about anatomic, functional, and metabolic changes [62]. The signals in MRI originate from stable nuclei with magnetic properties that can be measured through nuclear magnetic resonance.

We will mainly focus on the latter, since MRI methods have been extensively used in the context of computational oncology [40, 38, 64, 39]. Usual MRI methods, T1 and T2 sequences, also termed anatomical methods, allow to coarsely separate solid and fluid phases. The contrast in T1 method is given by the characteristic time it takes for water protons to synchronize, which depends on the surrounding tissue. This time is denoted inversion time T1. Fluid phase, like cerebrospinal fluid (CSF) in the brain, remains dark with the short time sequence of radio frequency/echo measurement of T1 method (T1 $\approx$ 200 ms), but becomes bright with a very long sequence T1 method (T1 $\approx$ 3000 ms), denoted FlAIR (standing for fluid attenuated inversion recovery). Conversely, the contrast of T2 method is given by the loss of synchronicity of protons (see [62], ch.4). The characteristic time T2 of synchronicity loss depends on the tissue type, for instance, T2 $\approx$ 100 ms for fat and T2 $\approx$ 2000 ms for CSF. The T1 method can be enhanced by the addition of the vascular network of a radiotracer such as Gadolinium, which remains bright under T1 sequence. This method is termed T1 contrast enhanced (T1-CE).

Another class of methods used in imaging-informed modeling is the dynamic contrast enhanced MRI (DCE-MRI). It has been firstly used to obtain information on peripheral perfusion, for vascular-related disease of brain, heart and muscular system [65]. In DCE-MRI, a Gadolinium-based contrast agent is injected into the vascular system and T1 sequences are acquired every few seconds before, during, and after the injection. Specifically on brain tumors, the relative cerebral blood volume (rCBV) method is a class of DCE-MRI and can be used as a prognosis* tool itself [66]. By this method, a leaky vasculature will be revealed. This is a fundamental feature in the case of brain cancer: the blood-brain barrier, strictly impervious to blood flow, with an abnormal permeability will indicate a potential tumorous neo-vasculature. From the modeling point of view, the rCBV method allows quantifying the permeability between intra- and extra-vascular space [67].

A third class of MRI is used to set the parameters of mathematical models: the diffusion-weighted MRI (DW-MRI). This method has been designed to measure the diffusion of water within tissue. This sequence corresponds to a magnetic field with a linearly varying intensity, which is prescribed successively in two opposite directions [68]. In short, the respective intensities of the magnetic field can be retrieved in both directions, and their spatial evolution characterizes the heterogeneous diffusion of water within the excited tissue. This quantity is expressed as the apparent diffusion coefficient (ADC) of water, and it is inversely correlated to cellular density [39]. In case of organized fibrous tissue, the water diffusion increases in the parallel direction of the fiber and decreases in the perpendicular direction [69], such that diffusion is then denoted as anisotropic. DW-MRI can be extended to the 3D case to render this anisotropy, leading to a method termed diffusion tensor imaging (DTI). In DTI-MRI, the diffusion is evaluated at the minimum of 6 non-colinear vectors - as the resulting tensor is symmetric - with preferably equal angles [68, 69]. DTI-MRI can be used to map the white matter tracts in the brain. However, this method is not used routinely: an acquisition lasts for more than an hour, and this duration could be an additional burden for the patient.

These mechanical quantities measured by MRI can be completed by measurements of the metabolism of key chemical quantities, such as oxygen and glucose. The quantification of oxygen levels, and therefore the hypoxia*, is a critical feature for cancer staging[70]. On a patient-specific basis, this quantification may hold importance in future modeling. The PET scan Cu-ATSM - standing for Cu-diacetyl-bis($N^4$-methylthiosemicarbazone) - has very promising *in vitro* results and is still in development for clinical applications [71]. Cu-ATSM is a radiotracer which accumulates in hypoxic tissue due to the chemical imbalance of low oxygenation. Among MRI methods, there is the blood oxygen level dependent method which can distinguish, by modified T2 sequence, oxygenated and non-oxygenated haemoglobin by their different magnetic properties. However, this method is more useful for acute hypoxia* than for chronic hypoxia*, characteristic of cancer [70]. To quantify the glucose metabolism of cancer, the PET radiotracer used is F-FDG. Analogous to glucose, it becomes radioactive by addition of the isotope fluor 18. As tumors have an important consumption of glucose, F-FDG is preferentially trapped by them, even if it does not have the energy properties of glucose. This method is routinely used for cancer diagnosis and treatment response [72].

*Segmentation.* This procedure consists of the delineation of different regions of interest in the imaged tissue and provides the anatomical input to define the geometry and spatial parameter maps in mathematical models of cancer growth and treatment response. Segmentation has been traditionally performed by radiologists and specialized physicians. However, manually performing this task can be difficult, time-consuming, and prone to high intra- and inter-observer variability. During the past two decades, several semi-automatic or automatic methods have been developed, such as hidden Markov chain in 2001 [73], supervised learning and convolutional neural network (CNN), which are still under development in 2020 [74]. CNN is now the leading method for semi-automatic and automatic segmentation. Some examples of its applications are DeepMedic in the brain [75] and 3D AxelNet in the prostate [74]. These methods are 3D CNN, which means that the weights within the layers of the neural network are not linked to a specific MRI slice, but directly influence the treatment of the entire 3D data space. Nevertheless, the training of CNN requires an important amount of data. A well-known example is the brain tumor segmentation (BraTS) dataset [76], used to benchmark the brain tumor segmentation. BraTS contains 542 imaging datasets of patients with high- and low-grade glioma*, composed of anatomical methods, contrast enhanced and FLAIR sequences. The imaging data are interpolated to a uniform 3D space with isotropic resolution, allowing for a 3D CNN implementation. The patient cohort is divided in three subgroups: the training set (242 patients), which is used to calibrate the weights of the CNN; the validation set (66 patients), which is used to assess the calibrated weights; and the testing set (191 patients), which is used to evaluate the accuracy of the CNN. Thus, the training of a CNN for MRI segmentation - *a fortiori* for the forecast of cancer evolution (see [77, 78]) -, requires an amount of homogeneous data which is not always available.

*2.2. Patient-specific approaches in clinical scenarios*

Computational oncology is becoming an increasingly rich and prominent field of cancer research [39, 38, 64]. Multiple patient-specific biomechanical models have been proposed to reproduce and forecast the growth and treatment response of different types of cancers in several clinically-relevant applications, e.g., in brain [79, 32, 60, 80], breast [31, 81], prostate [38, 30], kidney [82], and pancreatic tumors [83]. In this section, we proceed to describe in detail a small selection of some illustrative recent studies in breast, brain, and prostate cancer. The examples considered herein share the same framework of tumor growth inhibited by solid stress. Clinical imaging data from individual patients are used to initialize the model, estimate parameters, and assess model forecasts after calibration. The integration of clinical data within the mechanistic framework is briefly described. For a more detailed review of imaging types, clinical applications, modeling paradigms, and computational methods for patient-specific models of cancer growth and therapeutic response, the interested reader is referred to [38].

*Breast cancer.* Neoadjuvant therapy (NAT) is widely regarded as a standard-of-care treatment for locally advanced breast cancer before surgery [84]. Patients who achieve a pathological complete response (i.e., no residual disease) at the completion of NAT have a greater likelihood of recurrence-free survival [85]. Thus, the early determination of response to NAT would enable physicians to modify the therapeutic regimen of a non-responding patient (e.g., by changing dosage, dose schedule, prescribed drugs), and thereby improve treatment outcomes while avoiding unnecessary toxicity. To address this challenge, Jarrett *et al.* [31] propose to use personalized, computational forecasts of tumor response to NAT regimens obtained with a mechanistic model calibrated with patient-specific longitudinal quantitative MR data acquired early in the course of NAT.

In [31], Jarrett *et al.* describe breast cancer growth in terms of tumor cell density as a combination of tumor cell mobility and proliferation, which are modeled with a diffusion operator and a logistic term, respectively [86, 87, 88]. Additionally, Jarrett *et al.* introduce a sum of linear terms in the tumor dynamics equation to describe the NAT drug-induced death of tumor cells [88]. Each of these therapeutic response terms corresponds to one drug included in the patient's NAT regimen and is proportional to the local drug concentration in the breast tissue and the local tumor cell density. The initial spatial distribution of each NAT drug concentration in the breast tissue after infusion is estimated from DCE-MRI data and it follows an exponential decay in time, which reasonably approximates the relevant drug pharmacokinetics during the time scale of NAT. Additionally, the model presented by Jarrett *et al.* is coupled with quasistatic linear mechanics to account for the tumor mass effect. To model the forces exerted by the tumor, the authors use a pressure term that is proportional to the local tumor cell density [89, 87]. Then, to model the mechanical inhibition induced by the tumor mass effect, the authors constrain the tumor cell diffusion coefficient with an exponentially decaying function of the Von Mises stress [87]. Additionally, an earlier study of a simpler version of this breast cancer model by Weis *et al.* explored the differences in tumor forecasting when using a linear elastic versus a neo-Hookean nonlinear hyperelastic constitutive law, with which they implemented several highly incompressible cases [90]. Weis *et al.* reported differences in tumor volume and total tumor cellularity below 10% with respect to the linear elastic case even in a highly incompressible scenario. Assuming the same Poisson ratio in both constitutive laws, the differences in these quantities were below 0.5%. Thus, Weis *et al.* concluded that the error in assuming a linear elastic constitutive model is admissible for this tumor forecasting application, especially since it would be included within the uncertainty of MR measurements of tumor cell density [90, 91, 92].

Jarrett *et al.* [31] use longitudinal anatomic and quantitative MR imaging datasets to initialize, parameterize, and assess the predictions of their model for each patient during the course of NAT. To this end, the authors present a robust imaging preprocessing pipeline. In brief, they first rigidly register the DCE-MRI and DW-MRI images acquired within each scan session using *imregister* in MATLAB. Second, tumor regions of interest are segmented on postcontrast DCE-MRI data from each scan session by means of a fuzzy c-means-based clustering algorithm developed in MATLAB [93, 94]. In addition, ADC maps are extracted from DW-MRI data using standard methods [91, 87] and the DCE-MRI data are analyzed using the Kety-Tofts model [95] to estimate parameters characterizing the local vasculature. Third, the contrast-enhanced images obtained from DCE-MRI data from each scan session are longitudinally registered using a non-rigid registration algorithm with a tumor-volume preserving constraint [96]. The resulting deformation map is applied to the other MRI modalities. Finally, the authors calculate several imaging-based fields that are required to initialize and parameterize the model: (i) the registered contrast-enhanced images are used to manually contour the breast domain and to generate adipose and fibroglandular tissue masks by means of a k-means clustering algorithm, which enable the definition tissue-specific material properties [87]; (ii) the Kety-Tofts parameters obtained from DCE-MRI data are used to calculate a normalized blood volume spatial map, which is used to define the initial NAT drug concentration fields mentioned above; and (iii) tumor cell density maps at each scan date are calculated by leveraging an inverse linear relationship with the ADC maps extracted from DW-MRI data [86, 87].

While most patients in the cohort ($N = 21$) analyzed by Jarrett *et al.* in [31] received two NAT regimens, the authors center their analysis exclusively on the first one. Hence, Jarrett *et al.* use the tumor cell density maps extracted from two consecutive scans for model calibration: one shortly before the onset of NAT to set initial conditions, and another within the first cycle of NAT to set a model-data mismatch functional. Their calibration method consists of a Levenberg-Marquardt least-square algorithm implemented in MATLAB [97, 31, 87], which aims at minimizing the aforementioned functional and hence yield constant values of baseline (i.e., unconstrained) tumor cell diffusivity, NAT drug efficacy, and NAT drug temporal decay along with a spatial map of the tumor cell proliferation rate on a patient-specific basis. The rest of the model parameters are fixed from the literature or from MRI measurements [31]. The calibrated model is then initialized using the datasets from the second scan and solved in time until the date of a third scan after the conclusion of the first NAT regimen. At this time point, the authors assess their model predictions of tumor cell density by comparing them with the corresponding tumor cell density map extracted from the third DW-MRI dataset. During both calibration and forecasting, the model is solved by means of a fully explicit finite-difference scheme implemented in MATLAB [31, 97, 38].

In [31], Jarrett *et al.* show that the imaging-computational framework described above renders highly accurate predictions of breast cancer response to NAT. To illustrate this computational technology, Figure 2A shows an example of a patient-specific model forecast against corresponding imaging data at the same date. In particular, the authors report a significant correlation between model forecasts and imaging measurements of total tumor cellularity, tumor volume, and tumor longest axis in a subgroup of 13 patients exclusively receiving chemotherapy for their first NAT regimen, with corresponding concordance correlation coefficients above 0.90 ($p < 0.01$). Additionally, Jarrett *et al.* leverage their model to explore a clinically feasible strategy to optimize NAT chemotherapeutic regimens on a patient-specific basis *in silico*. First, they define a set of alternative NAT regimens by fixing the total dose prescribed in the standard treatment for each patient while varying the number of doses, their frequency, and the corresponding drug dosage. In particular, the authors consider four alternative NAT regimens by equally dividing each original dose in two, three, four, or daily fractions that are delivered more frequently than in the standard treatment assuming an evenly distributed schedule based on the original NAT plan. Then, the authors run a simulation of their calibrated model for each of these alternative regimens accordingly and assess their performance in controlling breast cancer growth at the conclusion of NAT for each patient. Jarrett *et al.* report that an additional $0 - 46\%$ reduction in total cellularity (median 17%) might have been achieved across the patient subgroup ($N = 13$) compared to their original chemotherapeutic regimen according to standard-of-care NAT protocols. Indeed, the authors also report that alternative NAT regimens rendering a superior control of the patient's breast tumor were also found to significantly outperform the standard regimens ($p < 0.001$), thereby suggesting that standard regimens may not be the most effective for every patient and calling for the personalization of NAT to optimize therapeutic outcomes.

*Brain cancer.* Tumors originating in the brain have received a large attention from the computational oncology community since the ability to early identify, or even predict, progression through patient-specific model forecasts would facilitate treatment personalization of these highly lethal diseases, which would ultimately translate into a major impact on survival, functional status, and quality of life [79, 32, 60, 80, 98, 89, 58, 99, 100, 101]. Here, we focus on a recent study by Hormuth *et al.* [80] on high-grade gliomas*. These brain malignancies exhibit a very aggressive behavior characterized as a highly heterogeneous and infiltrative disease, which ultimately influences the high recurrence rates in these tumors despite the advances in therapeutic strategies combining surgery followed by adjuvant radiotherapy plus chemotherapy [102]. Hormuth *et al.* posit that patient-specific computational forecasts of cancer growth and treatment response characterizing the unique underlying biology of the patient's tumor could enable the early identification of aggressive disease progression and the selection of more effective therapies on a personalized basis.

Hormuth *et al.* [80] present a family of 40 biologically-inspired, MR-informed models to describe the growth of high-grade gliomas* and their response to chemoradiation, which they analyze in a model selection framework to determine the model that best balances predictive accuracy and formulation complexity. Each model is constructed by choosing three key components in their formulation: (i) a biomechanical model describing tumor dynamics, featuring either one or two cancerous species; (ii) a global value versus a spatially varying map to define the tumor cell proliferation rate; and (iii) one out of ten alternative formulations for the tumor response to chemoradiation. The model featuring a single cancerous cell species describes the growth of high-grade gliomas* using the same reaction-diffusion paradigm as the breast cancer model [58, 98, 103]. Inspired by an earlier work by Gatenby *et al.* [104], the authors extend this reaction-diffusion paradigm to a two-species formulation including both the enhancing tumor fraction observed in post-contrast T1-weighted MR images as well as the non-enhancing fraction showing an hyperintense signal in T2-FLAIR MR images. These two fractions are modeled as two different tumor cell densities following reaction-diffusion dynamics and also featuring a biological competition mechanism within the logistic term that describes tumor cell proliferation. While several alternative biomechanical models have been considered in the literature to describe the mass effect induced by brain tumors [79, 32, 60], Hormuth *et al.* follow the same rationale as Jarrett *et al.* [31] in their breast cancer model. Hence, both the single and the two-species model are coupled with quasi-static linear mechanics and their diffusion coefficients are mechanically inhibited through the evaluation of the Von Mises stress. To model the tumor response to concomitant radiotherapy and chemotherapy, Hormuth *et al.* assume that either treatment induces an independent and immediate reduction in the tumor cell species in the model (i.e., they do not explicitly introduce any synergistic mechanism). Hence, the authors propose four alternative formulations of the survival fraction to either treatment: one depending on the local tumor cell density; two depending on the level of perfusion measured through the enhancement ratio; and another that is constant throughout the whole tumor. Then, they select 10 pairs from these four options to represent chemotherapeutic and radiotherapeutic effects in their model selection study.

Hormuth *et al.* initialize, calibrate, and assess the performance of their model alternatives by leveraging longitudinal MRI datasets for each patient (N=9) consisting of two anatomical data types (contrast-enhanced T1-weighted and T2-FLAIR images) and one quantitative MRI data type (DW-MRI). To prepare the imaging dataset for use within the model, the authors perform a rigid intravisit and intervisit registration by leveraging *imregister* in MATLAB. The enhancing and non-enhancing, T2-hyperintense tumor regions are segmented using a semi-automated approach on the anatomic MRI datasets, consisting of sequential thresholding and manual adjustments by experienced radiologists. The brain-skull interface is manually segmented on the T2-FLAIR images. Additionally, Hormuth *et al.* use a k-means clustering algorithm [105] to obtain masks of white and gray matter to define heterogeneous material properties in the brain tissue. The authors estimate the tumor cell density in the single-species model and that of the enhancing fraction of the two-species model from ADC maps using the same approach as in the breast cancer model by Jarrett *et al.* [31]. In the non-enhancing, T2-hyperintense tumor region, since the relationship between imaging properties and tumor cell density is less clear, the authors initialize its value to a constant both for the tumor cell density in the single-species model and that of the non-enhancing fraction in the two-species model, although they also note that alternative approaches have been proposed in the literature [58, 99]. Hormuth *et al.* employ a fully-explicit finite-difference scheme to solve their models and a Levenberg-Marquardt least-square algorithm to calibrate the relevant patient-specific parameters in each model, similarly to the breast cancer model above [97]. The rest of the parameters are either taken from the literature or fixed by the authors. Additionally, Hormuth *et al.* use the Akaike information criterion (AIC) [106] to select the best model balancing model-data agreement and complexity in terms of number of calibrated parameters.

Hormuth *et al.* carry out their modeling selection study in a cohort of 9 high-grade glioma* patients with unresected or partially resected tumors, whose individual imaging datasets consisted of a baseline MRI study (prior to the onset of chemoradiation) and two to three MRI studies at 1 to 2-month intervals after the conclusion of their chemoradiotherapeutic plan. They found that the best model across all patients was the two-species model, with spatially varying proliferation rates, and both chemotherapeutic and radiotherapeutic effects coupled to an estimate of perfusion obtained *via* the enhancement ratio. Hormuth *et al.* further explore the predictive accuracy of this model by considering two to all the datasets for each patient during calibration, and then forecasting the calibrated model until the date of the remaining datasets (if any). Their results show a good model-data agreement at the voxel level, while at the global level they report higher accuracy in terms of the Dice coefficient and tumor volume error in the enhancing region than in the non-enhancing region. Figure 2B shows an example of a model forecast along with the corresponding imaging data at the same date for a high-grade glioma* patient using the computational technology presented in this study. In particular, the authors show a strong agreement between measured and model predicted tumor volume and total tumor cell count within the enhancing region, with Kendall rank correlation coefficients above 0.94 and 0.79, respectively. Thus, Hormuth *et al.* conclude that their methodology is a promising strategy to integrate multimodal imaging data into personalized biomathematical models of high-grade glioma* in order to forecast spatiotemporal changes in tumor growth and response to therapy.

*Prostate cancer.* Benign prostatic hyperplasia (BPH) and prostate cancer are two common pathologies among ageing men [107]. Most new cases of prostate cancer are organ-confined adenocarcinomas, while BPH consists of the progressive enlargement of the central gland of the prostate and usually provokes bothersome urological symptoms [108, 30, 107]. Prostatic tumors originating in larger prostates tend to present more favorable pathological features [109, 110], but the fundamental mechanisms that explain this interaction between BPH and prostate cancer are largely unknown and much debated in the medical community. In [30], Lorenzo *et al.* hypothesize that this phenomenon may be caused by the mechanical inhibition induced by BPH over prostate cancer growth, which the authors explore computationally by leveraging a new patient-specific model accounting for the mechanical interplay between both diseases.

The model proposed by Lorenzo *et al.* in [30] relies on a phase-field formulation of prostate cancer growth, whereby a variable termed phase field identifies and describes the joint spatiotemporal evolution of both healthy and cancerous tissue [108, 30, 111]. In this study, tumor growth is driven by nutrient-dependent proliferation and apoptosis. The authors consider a generic nutrient that follows reaction-diffusion dynamics. This model also features a reaction-diffusion equation to describe the local production of a key biomarker in the clinical management of prostate cancer: the serum Prostate Specific Antigen (PSA) [108]. The integration of the local PSA field over the prostate volume ultimately yields the usual serum PSA values that physicians would consider in clinical decision-making. To represent the pathological deformation of the prostate due to tumor growth and BPH development, Lorenzo *et al.* follow a similar approach to the previous brain and breast cancer models [31, 80] that also accounts for the observed higher stiffness of the prostate central gland with respect to the surrounding peripheral zone [30]. The volumetric expansion induced by BPH is modeled with a time-dependent pressure term that is only defined within the central gland. Likewise, the tumor mass effect is modeled with another pressure term defined upon the phase field, thereby coupling tumor dynamics with the mechanical deformation of the prostate. To account for the confinement of the prostate within the pelvic region, the authors leverage Winkler-inspired boundary conditions over the external surface of the prostate. Additionally, Lorenzo *et al.* define a mechano-transductive function in the phase-field equation that progressively inhibits prostatic tumor growth as the local mechanical stress field intensifies. To this end, they use an exponentially decaying formulation of the stress level, similarly to the models discussed for breast and brain cancer above [31, 80] which, however, accounts for the surrounding tissue distortion and the usual internal stress state within growing tissue [18, 30].

Lorenzo *et al.* perform the study in [30] over the anatomy of the BPH-enlarged prostate of a patient with a tumor in the peripheral zone, which they extracted from MR imaging data available at the Initiative for Collaborative Computer Vision Benchmarking (I2CVB: i2cvb.github.io) [112]. For each patient with biopsy-confirmed PCa, this public database features a single-visit dataset including: T2-weighted MR images, DCE-MR images, DW-MR images, MR spectroscopic images, ADC maps, and the segmentations of the prostate, its local anatomy (i.e., central gland and peripheral zone), and the tumor delineated by an experienced radiologist [113]. Lorenzo *et al.* construct a volumetric Non-Uniform Rational B-spline (NURBS) mesh of the patient's prostate by leveraging a parametric mapping method [114, 115], whereby they deform the outer surface of a torus solid NURBS model to match with a surface model of the patient's prostate. To this end, Lorenzo *et al.* use 3DSlicer [116] to build a triangulated surface model of the patient's prostate by leveraging the T2-weighted images and the anatomic segmentations obtained from I2CVB, which they posteriorly smooth in Meshlab [117]. The resulting solid NURBS model of the patient's prostate is then refined by using standard knot insertion to ensure high accuracy in model simulations, which rely on isogeometric analysis (IGA) [118, 119]. Finally, the segmentations of the central gland and the patient's tumor are $L^2$-projected over the prostate mesh to define zone-dependent material properties and set the initial conditions of the tumor phase field. To define the BPH load and the parameter describing the Winkler-inspired boundary conditions, Lorenzo *et al.* run an inverse calculation that aims at deforming the patient's prostate back to a healthy state according to standard anatomical features [107]. This procedure is performed without considering the tumor. As a by-product, this calculation also renders a rough estimate of the mechanical stresses that BPH had induced in the patient's prostate before the diagnosis of prostate cancer at the MRI date, which is used as a prestress in the simulation study presented in their work [30]. The remainder of the model parameters are fixed from the literature or computationally estimated [108, 30].

Using the biomechanical approach described above, Lorenzo *et al.* qualitatively show that the accumulation of mechanical stress due to prolonged BPH in enlarged prostates hinders the growth of prostatic tumors, which tend to develop more slowly, with limited invasiveness, and maintaining a smaller volume for a longer time. Thus, the mechanism described in [30] would contribute to explaining the favorable pathological features observed in tumors originating in larger prostates [109, 110]. Figure 2C shows an example of a forecast of prostatic tumor growth over a patient-specific prostate mesh using the methodology proposed by Lorenzo et al in [30]. Additionally, the mechanism proposed by Lorenzo *et al.* suggests that medical therapies aiming at reducing the volume of the prostate to alleviate BPH symptoms could potentially reduce the mechanical constraint on the tumor, and ultimately favor its development. Indeed, the authors analyze this possibility in a posterior study [120], where they explore the effects on prostate cancer of a type of common BPH drugs known as $5\alpha$-reductase inhibitors (e.g., finasteride, dutasteride). These drugs may reduce the prostate volume by approximately 18% to 26% and can also promote apoptosis in the tumor [121, 122, 123, 124]. In [120], Lorenzo *et al.* extend their model to account for these two drug-induced mechanisms by introducing an additional time-dependent pressure term describing prostate shrinkage and by increasing the apoptotic rate in the tumor dynamics equation, respectively. In particular, the new pressure term is parameterized by leveraging population-based observations from two large clinical studies [121, 122]. Lorenzo *et al.* used the same patient data from I2CVB to conduct a simulation study that shows that the maintenance of a mechanical constraint on tumor growth ultimately depends on the long-term competition between the apoptotic boost, which inhibits tumor growth, and the decrease in mechanical stress caused by prostate shrinkage, which favors tumor growth. In particular, their simulations show that tumors experiencing a major stress relaxation from drug-induced prostate shrinkage and only a limited upregulation of apoptosis ultimately exhibit more aggressive dynamics. In [120], the authors further hypothesize that this unfavorable outcome described by their model may contribute to explaining the larger proportion of higher-risk tumors in the treatment arm of two critical clinical trials exploring the use of $5\alpha$-reductase inhibitors for chemoprevention of prostate cancer [125, 126], which remains an open medical debate.

---

**Tumor growth forecasting: key points**

- Clinical imaging data provide physical measurements that can be leveraged to define the tumor-bearing organ geometry, initialize a mathematical model, and calibrate its parameters on a patient-specific basis.

- Once calibrated, computer simulations provide personalized forecasts that can be assessed against further clinical imaging data collected at posterior dates.

- A general modeling framework is presented: solid mechanics, with mechanically-inhibited tumor growth, coupled with reaction-diffusion of chemical agents. Clinically-relevant applications of this framework are provided.

- We identify two key challenges: (1) assessing uncertainty in clinical imaging data and model outcomes, and (2) transfer these forecasting technologies to the clinic.

Table 2: **Representative studies of mechanically-constrained tumor growth and treatment response studies in the clinical setting.** For further studies and additional details on the underlying modeling framework and computational technologies, the interested reader is referred to [38, 64, 39]

| Authors | Framework | Location \| sub-types | Objective | Nb. patients |
|---|---|---|---|---|
| Jarrett et al. 2020 [31] | Mechanically-constrained reaction-diffusion dynamics to describe tumor growth and neo-adjuvant therapy (NAT) response. Tissue deformation is modeled with linear elasticity. | Breast | Forecast patient's response to NAT and explore treatment personalization | 21 |
| Hormuth et al. 2021 [80] | 40 models evaluated, composed of: mechanically-constrained reaction-diffusion dynamics to describe tumor growth and treatment response, tissue deformation modeled with linear elasticity, one or two tumor cell species, concomitant chemo- and radio-therapy. | Brain \| high-grade glioma* multiforme | Forecast tumor evolution after surgery during concomitant chemo- and radio-therapy | 9 |
| Lorenzo et al. 2019 [30] | Mechanically-constrained phase-field model to describe tumor growth and reaction-diffusion dynamics for a generic nutrient and the prostate-specific antigen. Tissue deformation is modeled with linear elasticity. | Prostate \| Concomitant prostate cancer and benign prostate hyperplasia (BPH) | Relationship between BPH and favorable prognosis* of prostate cancer | 1 |

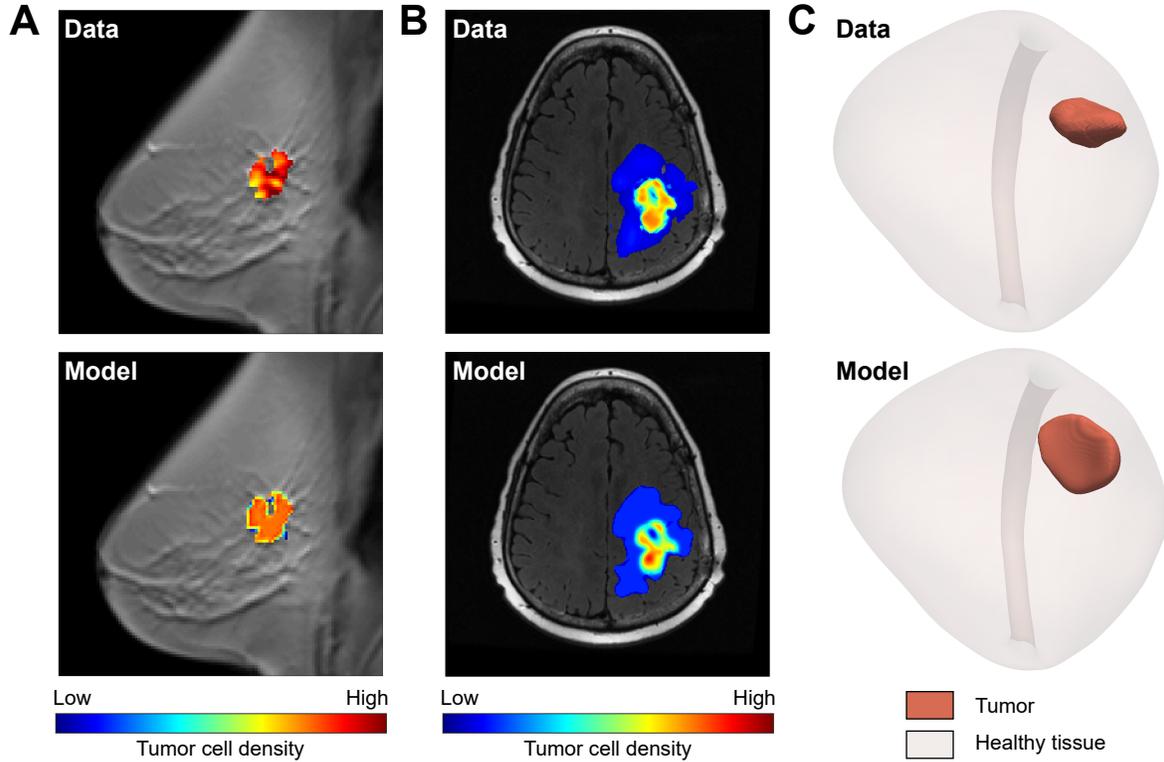

Figure 2: **Imaging-informed computational forecasting of tumor growth and treatment response in clinical settings.** Anatomic and quantitative clinical imaging data can be leveraged to initialize and parameterize mechanistic mathematical models of cancer growth and therapeutic response on a patient-specific basis [39, 40, 38]. The personalized model can then be used to run computer simulations to forecast the growth of the patient's tumor and the efficacy of treatments. These simulations can be further validated by comparison with additional imaging data from the patient at a posterior date. **Panel A** illustrates an application of this framework in the setting of NAT of breast cancer by Jarrett *et al* in 2020 [31]. The top figure shows a measurement of tumor cell density obtained from an ADC map calculated from DW-MRI data and plotted within the tumor region of interest identified on DCE-MRI data. The background anatomic image is a contrast-enhanced T1 map of the patient's breast. The bottom figure provides a model forecast of tumor cell density at the same location and time as the MR measurement shown above. **Panel B** illustrates an analogous ADC-based measurement (top) and model forecast (bottom) of tumor cell density at the same date during the post-surgical chemoradiotherapy of a glioblastoma* patient by Hormuth *et al* in 2021 [80]. In this case, the background anatomic data is a T2-FLAIR image of the patient's brain. **Panel C**, top figure shows a three-dimensional rendering of an isogeometric mesh of a patient-specific prostatic anatomy extracted from a T2-weighted MR image along with the L2-projected segmentation of an organ-confined tumor, which was obtained using the methods described in Lorenzo *et al* in 2019 [30]. The bottom figure shows a forecast of the tumor morphology at a later date using the mechanically-constrained model proposed in [30], in which the local mechanical deformation of prostatic tissue induced by concomitant BPH and prostate cancer exerts an inhibitory effect on tumor growth.

## 3. Clinical applications of mechanics in cancer management

Two classes of mechanical effects exposed *in vitro* transverse this review: i) tumor growth inhibition, which can help to design clinically-relevant modeling as shown in the previous section, ii) mechanically-induced phenotype switch *via* integrin signaling, which have been in-depth studied [127, 24, 128, 19] for more than twenty years. In the last years, the understanding of this phenotype switch led to new therapeutic targets. In 2020, it also led to an promising prognosis* tool [15]. We present these results in this section. We should note that, although stroma* mechano-biology constitutes a new reservoir of powerful tools, it is still underrepresented in the modeling community.

Mechanics is not only defined by the building of mathematical problems, it is also the underlying discipline enabling the design of new experimental setups. In particular, fluid mechanics underlies important applications in cancer management. Experimental setups for the isolation of circulating tumor cells are built upon microfluidic devices. The determination of the optimum diameter of a nanocarrier also requires fluid dynamics, as does the normalization of tumor vasculature induced by drugs. These two examples of clinical applications are presented into the fluid dynamics subsection.

Non-thermal electrophysics has already matured and journeyed from pre-clinical to clinical. A well-known application, the irreversible electroporation (IRE) is applied in clinics for fifteen years (see Davalos *el al.* [129]). As the *in vitro* findings date back to 1950, we decide to consider the application of electrophysics from the clinical side only. IRE applies on unresectable tumors and on poor prognosis* location such as the pancreas. We present section 3.3 the current developments of IRE, such as its coupling with immunotherapy. To conclude the clinical applications of mechanics in cancer, we present the idea of tumor-treating fields, a recent development in non-thermal electrophysics, the. Tumor-treating fields were approved in 2019 by US Federal Drug Agency as a new treatment for recurrent glioblastoma* multiforme.

*3.1. Mechano-biology*

We present the clinical applications derived from the mechano-biological findings presented in section 1.2. For a detailed description of these *in vitro* experiments, the reader will be referred to the section 1.2.

*In vivo mechano-biology of colorectal cancer.* Recently, Bauer *et al.* [15] have proven that stroma* mechano-biology and its consequences can be used for the prognosis* of metastasis in colorectal cancer. The cancer-associated fibroblasts* contribute to a stiffer and denser ECM network which surrounds the colorectal cancer[2]. This increasing mechanical stress environment will lead malignant carcinoma* cells to an epithelial* to mesenchymal* transition (EMT), *via* integrin downstream signaling (see section 1.2 for a detailed description). Disorganized carcinoma* cells after EMT have a metastatic profile. When the ECM becomes stiffer, the secretion of activin A by intestinal epithelial* cells increases. The level of activin A has been measured in 28 control and 28 patients of the Chicago Colorectal Cancer Consortium. It correlates with stage* IV metastatic cancers, the level of activin A being 4-fold as high as in the control group. Additionally, there is no significant difference in level between controls and all pre-metastatic stages* I-III.

---

[2]*Stricto sensu*, this phenomenon is termed fibrosis. However, in medical vocabulary, it could be reserved for scars and wound healing. More generally, and in the context of clinical oncology, this phenomenon will be rather termed desmoplasia*.

For decades, the integrin downstream signaling and its consequences on EMT have already been studied in-depth *in vitro* [24],[128] and with animal model [23], [26]. This direct clinical applications for metastatic prognosis* is one more landmark of the critical importance of the ECM mechano-biology in carcinoma* malignant evolution.

*Targeted drugs: focal adhesion* kinase inhibitor.* The mechanical signaling applied by the ECM is received by cell cytoskeleton through the transmembrane proteins, such as integrin. The cell in return tends to balance these mechanical stimuli by contractility and to maintain the equilibrium. In these conditions, the equilibrium is termed the tensional homeostasis. When the stiffness of ECM increases - hence the corresponding mechanical signaling - the magnitude of the cell mechanical response will increase accordingly. At the cell surface, this mechanical response is allowed by the clustering of integrin proteins, denoted focal adhesions* (FA). This mechanism is described section 1.2. These FA contain multiple mechanosensors, where focal adhesion* kinases (FAKs) play a specific role. FA allow increasing the generated traction force by actomyosin activity at the cell's surface. This instantaneous mechanical response is modified by FAKs. By altering the phenotype* of actomyosin, FAKs provoke a sustained tensional response. This augmented and sustained cellular tension will ultimately lead to enhanced tumor cell growth, survival, and invasion [130]. Lindsey *et al.* described in [47] the consequences of FAKs hyperactivity on well-organized epithelial* tissue. As the cell-matrix adhesions increase in number and intensity, the cell-cell adhesions essential to the organization of epithelial* tissue are lost. These cells then gain the ability to move individually, and this ability is characteristic of carcinoma* after EMT. Considering these mechano-biological findings, the inhibition of FAKs emerge as a potential therapeutic target.

In 2016, Jiang *et al.* [26] design a pre-clinical study in mouse models of FAKs inhibitor in pancreatic ductal carcinoma* (PDAC). This disease has a specific desmoplastic* stroma*, that acts as a barrier to T lymphocytes infiltration. At the tumor cell's level, this abundant and stiffer stroma is sustained by FAKs hyperactivity. Two groups of mouse models are treated, one with a microscopic early stage* PDAC and the other with a palpable late stage*. The prescription of a single-agent FAKs inhibition (VS-4718 in this study) significantly limits the tumor progression and increases responsiveness to immunotherapy PD-1 antagonists. VS-4718 leads to a significant extension of survival in both early and late treatment groups (4 and 8 months respectively). Promising results in animal models then authorize clinical trials with patients.

In 2020, Mohanty *et al.* reviewed in [56] the progress of FAKs inhibitors in clinical trials. Among these inhibitors, only 4 (PF-00562271, GSK2256098, VS-6063, and BI 853520) moved to subsequent development and were tested in clinical trials in solid tumors. These trials demonstrate the cytostatic activity of FAKs inhibitors. In other words, FAKs inhibitors stop the disease progression, without reducing the tumor volume. They extend the progression-free survival of 12 weeks in severe advanced solid tumors, as mesothelioma or PDAC. This extension is not accompanied by a clinical response, that is to say, the disease remains stable during this extension. The trials also show that the patients response could be correlated with two biomarkers, E-cadherin and merlin. Concerning the E-cadherin relationship with FA, the reader is referred to the section 1.2. In these trials, FAKs inhibitors are currently tested in several combinations of chemotherapy and immunotherapy.

*3.2. Microfluidics of metastasis and tumor vasculature*

The interplay between circulating fluids and cancer are numerous and his large topic would require a review by itself. The applications of this field in cancer management in particularly inspiring. Indeed, *in vitro* findings, modeling framework and adaptation to biological cues work together towards clinical applications.

We selected two paradigmatic examples of fluid mechanics application on cancer management. Over the last years, nanoparticle delivery has emerged as an important topic in which biomechanical models can make an important contribution. The seminal review of Jain in 2005 [131] pointed out the critical aspect of the normalization of the tumor blood vessels. The first example is a direct application of this principle: the injection of blocking vascular endothelial growth factor will normalize the poorly perfused tumor vasculature, resulting in an increase in pressure gradient, and in a reduction of the vessels diameter. These coupled phenomena can be advantageously modeled and exploited by fluid mechanics, to optimally design the nanocarriers diameter.

The second example belongs to the experimental field of cancer diagnosis of metastasis: the circulating tumor cells can be filtered or captured by microfluidic designs. Once again, complementary biological and mechanical approaches lead to successful applications. They overcome the sparsity of metastatic cells in blood, as well as heterogeneity.

The physical aspects of fluid mechanics in cancer are reviewed in [132]. From a clinical point of view, fluid mechanics' applications in cancer are reviewed in [16].

*Anti-angiogenic therapy and nanoparticles delivery.* In [133], Chauhan *et al.* study the effect of anti-angiogenic treatment on the delivery of nanoparticles. The drug is DC101 (ImClone System), a blocking vascular endothelial growth factor (VEGF) receptor-2, and two nanoparticles are tested: Abraxane, 10 nm diameter and Doxil 100 nm diameter. The experiments are performed on severe combined immunodeficient mice. DC101 and one of the nanoparticles diameter are injected on days 0, 3 and 6 and imaging by intravital microscopy are performed on days 2, 5 and 8 (imaging is performed through a window earlier implanted into the animal by surgery). Images are analysed by a custom code in Matlab, developed in a previous work of Chauhan *et al.* [134]. The experimental results give that DC101 injections decrease vessels diameters and enhanced between 2.7 to 3-fold the transvascular flux for the 10 nm particles, but with no change for 100 nm particles.

The mathematical model assumes classic Poiseuille's law for the blood flow, Starling's law for the intra/extra-vascular exchanges and the interstitial fluid flow follows poromechanics Darcy's law. The nanoparticles delivery is modeled by advection inside vessels and by advection-diffusion in extra-vascular space. The mathematical model shows that reducing vessel wall pore size decreases interstitial fluid pressure in tumors, allowing nanoparticles to enter more easily, but smaller pores associated with better hydrodynamics obstruct the penetration of bigger nanoparticles. The authors of [133] suggest that nanoparticles with diameters smaller than 12 nm are ideal for cancer therapy. These numerical results emphasize both the significance of the normalization of tumor blood vessels [131] and the efficiency of biomechanical modeling in the nanoparticles delivery challenge.

*Microfluidic and circulating tumor cells (CTC).* The isolation and characterization of CTC is a key challenge in metastatic cancer management. This can be achieved by the exploitation of biological properties (CTC biomarkers) or physical properties (size, deformation, viscosity). The first difficulty is the sparsity of CTC cells, which represent 0.004% of all cells in metastatic patient blood. Moreover, CTC present a high heterogeneity in biological and physical properties, thus, a microfluidic device solely based on size filtration and isolation could not be relevant as in a single patient CTC sizes vary between 4 and 30 $\mu$m. The review, in [135], reports several successful examples of dual approaches, biological and mechanical, which efficiently capture CTC heterogeneities.

In [136], Nagrath *et al.* present the CTC-chip device. Their approach is based on microposts coated with the antibodies EpCAM, standing for epithelial* cell adhesion molecule. The use of these specific antibodies for the CTC capture is justified by the over-expression of EpCAM by many carcinomas* (lung, colorectal, breast, prostate, among others) and by its absence of expression in haematologic cells. From the mechanical aspect, two parameters are critical for the CTC capture: the flow velocity, which determines the duration of the cell/micropost contact, and the minimization of the shear force, to ensure cell-micropost attachment. Additionally, these two parameters are constrained by a reasonable duration of patient blood analysis. The CTC-chip has been designed by simulation results. These results indicated that an equilateral triangle surface of 970 mm$^2$, with an architecture of 50 $\mu$m microposts separated by a shift of 50 $\mu$m every 3 rows, and a flow rate of 1 to 2 ml $\cdot$ h$^{-1}$ ensured the best compromise. The CTC-chip successfully identified CTCs in the peripheral blood of patients with metastatic lung, prostate, pancreatic, breast and colon cancer in 115 of 116 (99%) samples.

Three years later, Stott *et al.* in [137] proposed an improved design of the CTC-chip. Adding ridges on the wall of the device creates passive micro-vortex in the flow which notably increase the collision between CTC and EpCAM location. Therefore, the complex CTC-chip architecture could be simplified in superimposed micro-channels with their walls shaped by asymmetrical chevrons. The device is denoted Herringbone-Chip. Its design allows for 93% CTC captures at a flow rate 4-fold higher than CTC-chip. Its high-throughput and easier fabrication make it suitable for clinical use.

These two examples show that the collaboration between biological and mechanical approaches, with a well-designed modeling, lead to successful clinical applications. Experimental design of isolation of metastatic cells or modeling of nanocarriers is of critical importance for the future of cancer management. These are inspiring examples of the involvement of mechanics in cancer.

*3.3. Non-thermal electrophysics*

We conclude this review of the exploration of electrical treatments currently used in clinical management of cancer. The first one targets cell membranes: the irreversible electroporation (IRE). IRE is a non-thermal, minimally invasive technique which provokes tumor cell death by short high-voltage electric pulses. Since 2005 (see Davalos *el al.* [129]), IRE is a well-established treatment for non-resectable tumors. We focus on recent development of this technique on several locally advanced cancers. The second one targets cell mytotic activity: the tumor treating fields (TTF). TTF are the application of intermediate frequency alternating electric fields with low intensity. In 2019, TTF has been approved by the US Federal Drug Agency as a new treatment for recurrent glioblastoma* multiforme [138].

*Electroporation.* The Irreversible Electroporation is a non-thermal ablation technique which uses electric pulses to disrupt cell membrane [139]. Since 2005, more than 50 clinical trials have been applied to pancreatic, prostate, liver and kidney cancer showing that IRE has varying degree of clinical success, and reducing the progression of cancer [139]. Additionally, recent studies in human and mice suggest that IRE increases immune response of tumor-protective CD8+T cells. The success combination of IRE and immunomodulatory therapy (GC-CSF/STING agonist) shows significant improvement of treatment. In Shao *et al.* [140], an *in vitro* study shows that IRE enhances tumor antigen release. In the recent works of Burbach *et al.* [141], the hypothesis that primary treatment with IRE and following treatment with checkpoint immunotherapy increase the effective destruction and promote tumor clearance. In this study, authors also indicate that anti-CTLA-4 is a potential in-situ tumor vaccination strategy which could generate protective tumor-specific CD8+T cells, which play a critical role in host immune response.

In [29], Lin et al. conduct a clinical trial on the combination of IRE and immunotherapy by $V\gamma 9V\delta 2$ cells. 62 patients (32 control) with stage* III PDAC receive IRE therapy and 30 patients receive at least two cycles of $V\gamma 9V\delta 2$ cells infusion. This is the first reported immune treatment of PDAC by $\gamma$ T cells. Advanced PDAC is characterized by a strong immunosuppressive stroma*, IRE can improve, by its damaging effect, the drug delivery through this stroma*, but also increase tumor cells infiltration. Additionally, IRE provokes systemic change in tumor microenvironment, specifically in tumor immunity [142]. The combination of IRE and immunotherapy shows a significant improvement median survival (14.5 months, control 11). This is a promising strategy for treatment of stage* III PDAC.

*Tumor treating fields (TTF).* Glioblastoma* is a common brain tumor, which is very aggressive and recurrent. Its 5-year survival rate is still around 5% percent since 2000 [143]. Its standard of care is, if possible, maximum surgical resection followed by 6 one-week cycles a concomitant radiotherapy, temozolomide (TMZ) chemotherapy [50]. The TZM is used for its radio sensitizer effect, and after the 6 cycles, as chemotherapy maintenance for 6 or 12 months. Since 2005, no significance improvement has been made in this standard of care.

TTF [144] brought progress in glioblastoma* survival as shown in the phase III of clinical trial [145]. TTF are an antimitotic treatment, *i.e.* they affect dividing cells by delivering low-intensity alternating electric fields *via* electrodes arrays placed on the scalp. It is shown in [144] that the intermediate frequency of 200 kHz has the stronger effect on glioblastoma* cells.

In a phase III clinical trial, 637 patients, after the concomitant radiotherapy TZM treatment, receive TTF-TMZ maintenance or TMZ alone during 28 cycles. The TTF treatment shows a significant influence on progression-free (6.7 months *vs.* 4.0) and survival (20.9 months *vs.* 16.0). This treatment is a notable improvement on a particularly difficult cancer location. Recently, TTF has been approved by the US Federal Drug Agency as a new treatment for recurrent glioblastoma* multiforme [138].

---

**Clinical application: key points**

- Stroma* mecano-biology emerges as a promising field for prognosis* tools and new therapeutic targets.

- Fluid mechanics opened new opportunities to experimental design and nanocarriers design, of critical importance in the field.

- Non-thermal electrophysics is a well-established protocol for non-resectable tumors and its recent development led to promising new tools for pancreatic cancer and glioblastoma* multiforme.

## 4. Challenges and perspectives

In 2009, the National Cancer Institute deputy director Anna Barker wrote "we had reached an inflection point where we knew enough about the biology to bring in other fields". In 2012, Jennie Dusheck wrote a letter published in Nature, addressed to biologists and entitled *"Oncology: Getting Physical"* [146]. She pointed out that since the 1950's, age-adjusted cancer mortality rates have declined by only 11% whereas the budget allowed to treatment research have a 25-fold increase. This letter was mainly motivated by the examples given in the 2011 Nature review cancer of Mauro Ferrari *et al.* *"What does physics have to do with cancer"* [147], who has an important contribution in cancer nanomedicine [148] and cancer biophysical modeling [149, 34].

The years following this 'inflection point' have shown major reviews of cancer mechanical interplay as *"The role of mechanical forces in tumor growth and therapy"* in 2014 by Jain *et al.* [18]. The progress of clinical imaging led to the seminal review by T. Yankeelov *et al.* *"Clinically relevant modeling of tumor growth and treatment response"* in Science Translational Medicine in 2013 [39]. These studies showed decades of persevering research bringing to light and mechanics were finally allowed to 'make its part' in cancer research. Two examples to illustrate this statement: i) early research on elevated interstitial fluid pressure inside tumors as impediments of drug delivery in 1988 by Jain *et al.* [150] led to nanocarrier design through normalization of tumor vasculature in 2012 [133], ii) early work of Weaver *et al.* in 1997 [127] on integrin-mediated phenotype switch led to a clinical tool for early diagnosis of colorectal cancer with a metastatic prognosis* by Bauer *et al.* in 2020 [15]. Thus, the collaboration between biological and mechanical approaches could be a key for the next step of cancer treatments.

We have seen enormous efforts in the literature to select the best model, agent-based, individual-based, or focusing on partial differential equations, to best reproduce the phenomena of interest for computational oncology. However, the question of verifying whether the problem is solved to a satisfactory accuracy level is mostly neglected in literature. Recent efforts have been expended in the area of quality control and estimation for biomechanics problems, with the first real-time error estimation for surgical simulation proposed by Bui *et al.* in [151]. Additionally, adapting the computational mesh to obtain optimal accuracy in solving a model with minimal computational cost is a key strategy to obtain an efficient and precise implementation of computational oncology problems. Therefore, one of the most challenging problems in computational oncology resides in selecting the right model and identifying the best statistical distribution for each of its parameters (see Rappel *et al.* in [152]).

Personalized medicine has been a challenge for the last decade. Since 2017, numerous articles with promis-ing results are converging within the same modeling framework: i) the coupling of chemical reaction-diffusion equations and multi-physics mechanical equations, ii) the use of rich clinical patient-specific datasets with the robust experience of mechanical engineering. This framework is denoted as imaging-informed mechanical-based modeling. Yet the challenge of developing personalized medicine based on this framework is not achieved, we hope that the clinician will be willing to get acquainted with these promising new tools.

**Acknowledgments**

G.L. acknowledges funding from the European Union's Horizon 2020 research and innovation program under the Marie Skłodowska-Curie grant agreement No. 838786. D.B. acknowledges funding from the European Union's Horizon 2020 research and innovation programme under MSCA ITN ElectroPros No. 813192. The authors gratefully thank Dr. Angela M. Jarrett, Dr. David A. Hormuth II, and Dr. Thomas E. Yankeelov for providing the subfigures for breast and brain cancer shown in Figure 2 of this work.

# 5. Supplementary information

*5.1. Presentation of the modeling framework*

*Tumor growth approaches.*

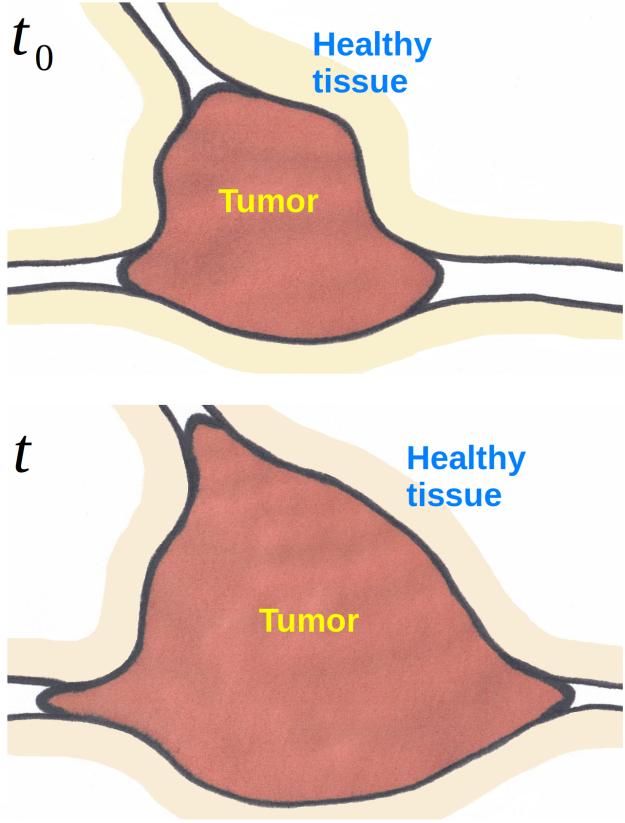

Figure 3: **Tumor growth approaches**
Phenomenological:

- Gompertz law:

$$N(t) = N_0 \exp\left[\ln\left(\frac{\theta}{N_0}\right)(1 - \exp(-\gamma t))\right]$$

$t$ time, $N(t)$ tumor cell density at time $t$, $N_0$ tumor cells initial density, $\theta$ carrying capacity of the domain, $\gamma$ growth rate (see [153]).

- Diffusion law: $\partial_t N = \nabla \cdot (D_T \nabla N)$
  $D_T$ the diffusion coefficient of tumor cells.

Mass conservation:

$$\begin{cases} \partial_t(\rho^T V^T) = \sum \stackrel{i \to T}{M} \\ \sum \partial_t(\rho^i V^i) = -\sum \stackrel{i \to T}{M} \quad i \neq T \end{cases}$$

$V^T$ tumor volume, $i$ other components of the system (for instance, interstitial fluid, stroma*). $\stackrel{i \to T}{M}$ mass exchange between phases $i$ and $T$.

In general, the tumor growth between $t_0$ and $t$ may be described by two approaches: the phenomenological approach and the balance law approach. The older and most famous phenomenological approach is the Gompertz law. The formulation comes from the work of Anna Laird in 1964 [153]. This adaptation of the sigmoid function can give good results *in vitro*, but the *in vivo* dynamics can not be properly reproduced. The Gompertz law may be replaced by a more general Logistic law: $\partial_t N = \gamma(x) N(t) \left(1 - \frac{N(t)}{\theta}\right)$ where the growth rate $\gamma(x)$ is spatially resolved. These laws can be replaced or completed by a diffusion law (see the models presented in section 2.2), where the tumor is considered as chemical species diffusing in the domain. The mass conservation approach poses the problem of cancer growth and treatment relying on fundamental physical balances (for instance, see [154, 35]). The mass of the system being conserved, the evolution of the tumor volume will be balanced by the evolution of the other components of the system (fluids, nutrients, and cells). Nevertheless, the mass exchange between the components $\stackrel{i \to T}{M}$ may be defined by various constitutive equations, physical or phenomenological.

*Tumor growth inhibition.*

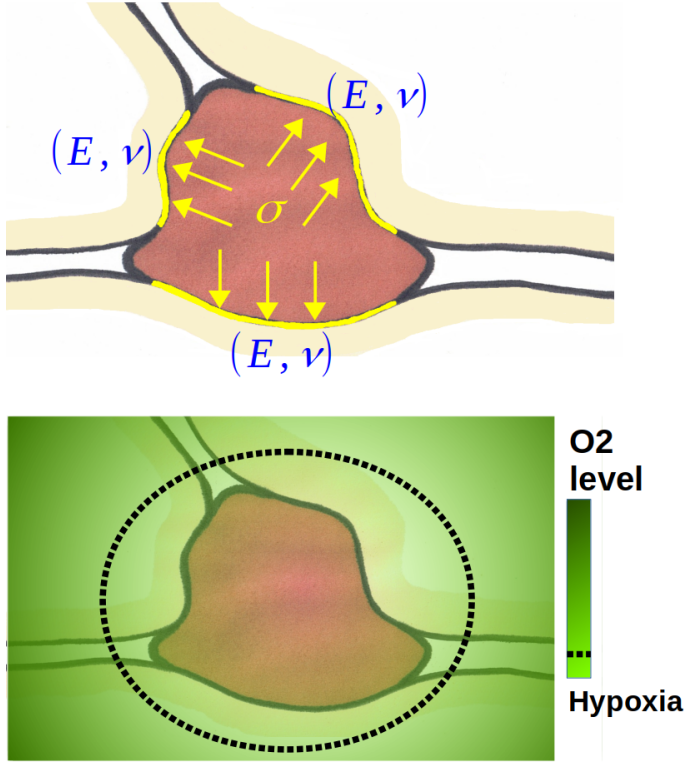

Figure 4: **Tumor growth inhibition** may be modeled as:

- Mechanical, with a stress inhibition threshold $\sigma_y$.

$$\bar{\bar{\sigma}} = \bar{\bar{\bar{C}}}(E,\nu) : \epsilon(\mathbf{u})$$

$$\sigma_{\text{VM}} = \sqrt{\frac{3}{2}(\sigma - \frac{tr(\bar{\bar{\sigma}})}{3}\bar{\bar{\mathbf{I}}})}$$

and test if $\sigma_{\text{VM}} \geq \sigma_y$

- Chemical, with a hypoxia* threshold $O_{2\text{crit}}$.

$$\partial_t(C) - \nabla \cdot (D_C \nabla(C)) = T_{\text{Source}} - T_{\text{Sink}}$$

and test if $C(t) \leq O_{2\text{crit}}$

where $D_C$ is the diffusion coefficient of the evaluated chemical agent, here oxygen.

We distinguish two main ways to model the tumor growth inhibition: mechanical inhibition through a stress criterion and chemical inhibition through a concentration threshold of a chemical agent.

From the displacement $\mathbf{u}$ and the material parameters, we can deduce the stress tensor $\bar{\bar{\sigma}}$. In figure 4, $\sigma$ is modeled with a linear elastic law. Then, a scalar stress metric, such as the Von Mises stress $\sigma_{\text{VM}}$, can be compared to an inhibition threshold $\sigma_y$. This stress could be compared to an inhibition threshold $\sigma_y$ (see figure 4).

Chemical inhibition of tumor growth may be modeled with various criteria (*e.g.* nutrients, treatment), in figure 4, it is oxygen. Oxygen level may be evaluated with a reaction-diffusion equation, as in figure 4, or by a heuristic relationship. Then it is compared with the hypoxia* threshold.

Depending on how the inhibition is modeled, dependencies will be added to the tumor growth model. The growth rate $\gamma$ may become $\gamma(O_2)$, the tumor cells diffusion coefficient $D_T$ may become $D_T(\sigma_{\text{VM}})$ or the mass exchange between the system components $\overset{i \to T}{M}$ may become $\overset{i \to T}{M}(O_2, \sigma_{\text{VM}})$.

*Specificity of poromechanics.*

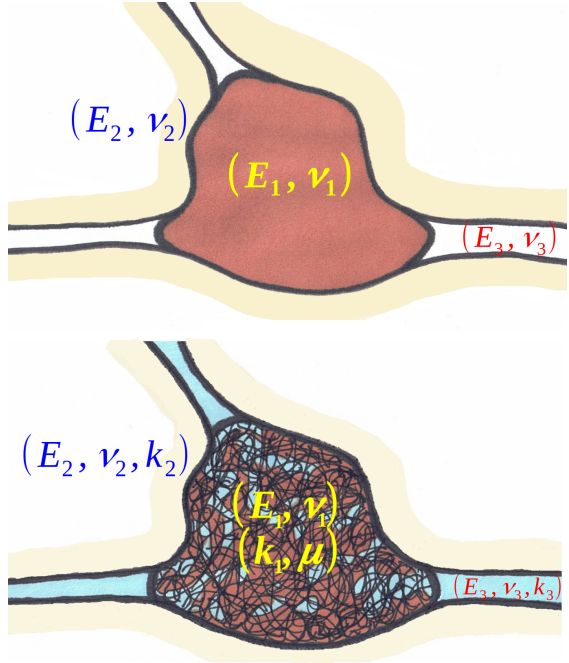

**Elastic solid and porous system mechanics**.

- An elastic solid is characterized by two material parameters, the Young's modulus $E$ -its stiffness- and the Poisson's ratio $\nu$ -its compressibility-. Each elastic solid in the domain has its own tuple $(E,\nu)$. The unknown of the system is the displacement field $\mathbf{u}$. The linear momentum balance, in absence of external load, reads:

$$\nabla \cdot \sigma = \mathbf{0} \quad \text{with}$$

$$\sigma = \frac{E\nu}{(1+\nu)(1-2\nu)}\epsilon(\mathbf{u}) + \frac{E}{2(1+\nu)}tr(\epsilon(\mathbf{u}))\bar{\bar{I}}$$

- A porous system is composed of porous solid(s) with inner fluid(s) passing through. An elastic porous solid is defined by the elastic tuple $(E,\nu)$ and its permeability $k$. The fluid is characterized by its dynamic viscosity $\mu$. The unknowns of the system are the displacement field $\mathbf{u}$ and the fluid pressure $p$. The system linear momentum balance, in absence of external load, reads:

$$\nabla \cdot \sigma^T = \mathbf{0} \quad \text{with} \quad \sigma^T = \sigma - p\bar{\bar{I}}$$

**Diffusion of chemical species**

- In an elastic solid a diffusive chemical species may be added. Its diffusion will depend only on its diffusion coefficient $D$. However, this coefficient may be space dependent $D(x)$ and partially coupled to the elastic solid through a stress inhibition threshold $D(\sigma_{\text{VM}})$.

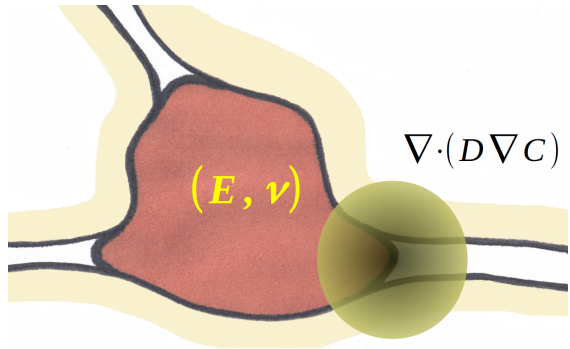

- In a porous medium, diffusive chemical species will also be advected by the inner fluid. A Darcy's term, dependent on both permeability $k$ and dynamic viscosity $\mu$, will be added to the diffusion equation.

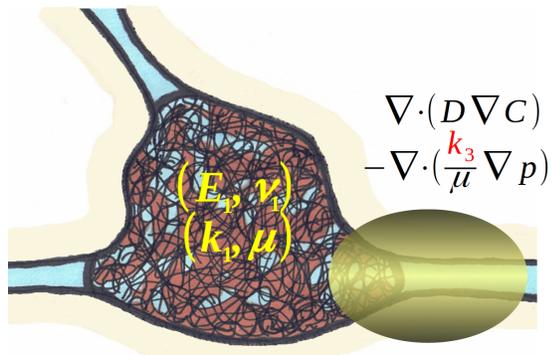

*5.2. Glossary of clinical terms*

**Adenocarcinoma:**  From Greek *Adeno* 'gland'. Subtype of carcinoma* which originates or presents the characteristics of a gland.

**Apoptosis:**  From Greek 'falling off'. Programmed cell death. Contrary to necrosis*, this is a highly regulated and controlled process.

**Carcinoma:**  From Greek 'crabe'. Subtype of cancer characterized by abnormal and excessive growth of epithelial* cells.

**Desmoplasia:**  Formation of connective tissue due to proliferation of fibroblasts*.

**Epithelial cells:**  These cells, thanks to their organized cell-cell adhesions, form a thin and continuous layer of packed cells that covers the outer surfaces of the organs. Epithelial cells also constitute the main component of glands.

**Etiology:**  Causes of disease.

**Extra-cellular matrix:**  Non-cellular material secreted by cells into the surrounding medium, which provide cell adhesion and intercellular communication. Depending on the cells, this material is mainly composed of collagen (fibrous structure), elastin (elastic property), fibronectin (cell adhesion) or glycosaminoglycans (shock absorption).

**Fibroblast:**  Mobile cell that produces the extra-cellular matrix* collagen.

**Focal adhesion:**  Sub-cellular structure that serves as the mechanical linkage to the extra-cellular matrix*, *i.e.* cell-matrix adhesion. This mechanical linkage forms by the concentration of integrin proteins (and other signaling proteins) upon fibronectin and other cell-matrix adhesion sites.

**Glioma:** Brain tumor originating of glial cells, mainly from oligodendrocytes or astrocytes.

**Glioblastoma:** Stage* IV glioma*. Aggressive and recurrent, characterized by necrotic core and/or micro-vascular abnormal proliferation. Glioblastoma almost exclusively come from astrocytes (rare hybrids may have an oligodendrocytes-astrocytes profile).

**Histology:** From Greek 'Tissue knowledge'. The study of the microscopic anatomy of tissues. Various processes are involved: light microscopy, cell staining or freezing.

**Hypoxia:** At the cell's scale, it refers to a state in which oxygen supply is insufficient. If this insufficiency is severe and prolonged, it could lead to necrosis*.

**Lymphoma:** Cancers of lymphocytes (white blood cells). The common feature of different types of lymphoma is the enlargement and the swelling of lymph nodes.

**Mesenchymal cell:** Undifferentiated cell, highly mobile, which can form various tissues (muscle, bone, neural tube). Epithelial* cells can lose their organization and return to the state of mesenchymal. This process is termed epithelial to mesenchymal transition (EMT).

**Necrosis:** From Greek 'death'. External factors, such as infection or trauma, cause cell deregulation leading to death, which provokes an inflammatory response of the surrounding tissue.

**Neoadjuvant therapy:** Therapy or a group of therapies (radio-, immuno- or chemotherapy) that are delivered before surgery or another primary treatment with curative intent.

**Parenchyma:** Functional part of an organ, complemented by the stroma*.

**Phenotype:** Observable characteristics resulting from the interaction of a genotype with its environment.

**Prognosis:** From Greek 'foreseeing'. Expected development of a disease.

**Stage:** Staging of cancer: I - local, benign; II invasion of nearby tissue; III lymph node invasion; IV metastatic, recurrent.

**Stroma:** Structural part of an organ, also termed connective tissue, complemented by the parenchyma*.

## 5.3. List of acronyms

| | |
|---|---|
| **ADC** | Apparent diffusion coefficient |
| **BOLD** | Blood oxygen level dependent |
| **CE** | Contrast enhanced |
| **CFD** | Computational fluid dynamics |
| **CNN** | Convolutional neural network |
| **CSF** | Cerebro-spinal fluid |
| **CT** | Computed tomography |
| **CTC** | Circulating tumor cell |
| **DCE-MRI** | Dynamic contrast enhanced magnetic resonance imaging |
| **DTI** | Diffusion tensor imaging |
| **DW-MRI** | Diffusion weighted magnetic resonance imaging |
| **ECM** | Extra-cellular matrix* |
| **EGF** | Epidermal growth factor |
| **EMT** | Epithelial* to mesenchymal* transition |
| **F-FDG** | $^{18}$fluor-fluorodeoxyglucose (radio-tracer) |
| **FAKs** | Focal adhesion* kinases |
| **FlAIR** | Fluid attenuated inversion recovery |
| **GBM** | Glioblastoma* multiforme |
| **H&E** | Hematoxylin and eosin |
| **HER2** | Human epidermal growth factor receptor 2 |
| **IRE** | Irreversible electroporation |
| **MEC** | Mammalian epithelial* cell |
| **MRI** | Magnetic resonance imaging |
| **PET** | Proton emission tomography |
| **rCVB** | Relative cerebral blood volume |
| **TTF** | Tumor-treating fields |